\documentclass[useAMS]{emulateapj}  

\usepackage{natbib}
\usepackage{amssymb}
\usepackage{amsmath}
\usepackage{amstext}
\usepackage{xcolor}
\usepackage[varg]{txfonts}	
\usepackage{microtype}		
\usepackage{multirow}

\let\url\relax
\makeatletter
\renewcommand{\fps@figure}{tp}
\makeatother
\setkeys{Gin}{width=\linewidth,height=0.9\textheight,keepaspectratio=true}

\newcommand\hbeta{\ensuremath{\mathrm{H}\beta}}
\newcommand\halpha{\ensuremath{\mathrm{H}\alpha}}
\newcommand\elec{\ensuremath{_{\mathrm{e}}}}

\newcommand{\te}{\ensuremath{T\elec}}
\newcommand{\nel}{\ensuremath{n\elec}}

\newcommand{\vlsr}{\ensuremath{V_{\mathrm{LSR}}}}
\newcommand{\chb}{\ensuremath{c(\hbeta)}}

\newcommand\unit[1]{\ensuremath{\mathrm{#1}}}
\newcommand\K{\unit{K}}
\newcommand\cmc{\unit{cm^{-3}}}
\newcommand\kms{\unit{km\ s^{-1}}}

\newcommand\wav[1]{\ensuremath{\lambda #1}}
\newcommand\wavs[2]{\ensuremath{\lambda\lambda #1,#2}}

\newcommand\hst{{\it HST}}

\slugcomment{The Astrophysical Journal, XXX:XXXX--XXXX, 2014 XX XX}

\shorttitle{CNO trace in NGC~6888}
\shortauthors{Mesa-Delgado et al.}

\begin{document}


\title{The Trace of the CNO Cycle in the Ring Nebula NGC~6888\altaffilmark{*}}


\author{A. Mesa-Delgado$^1\dagger$}
\affil{$^1$Instituto de Astrof\'isica, 
  Facultad de F\'isica, Pontificia Universidad Cat\'olica de 
  Chile, Av.~Vicu\~na Mackenna 4860, 782-0436 Macul, Santiago, Chile}
\email{$\dagger$E-mail: amesad@astro.puc.cl}

\author{C. Esteban$^{2, 3}$, J. Garc\'ia-Rojas$^{2, 3}$,}
\affil{$^2$Instituto de Astrof\'isica de Canarias, E-38200 La Laguna, Tenerife, Spain}
\affil{$^3$Departamento de Astrof\'isica. Universidad de La Laguna, E38205 La Laguna, Tenerife, Spain}

\author{J. Reyes-P\'erez$^{4}$, C. Morisset$^4$ }
\affil{$^4$Instituto de Astronom\'ia, Universidad Nacional Aut\'onoma de M\'exico, Apdo. Postal 70264, M\'ex. D. F., 04510 M\'exico}

\author{and F. Bresolin$^5$ }
\affil{$^5$Institute for Astronomy, 2680 Woodlawn Drive, Honolulu, HI 96822, USA}




\altaffiltext{*}{Based on data collected at Subaru Telescope, which is operated by the National Astronomical Observatory of Japan.}

\begin{abstract}
We present new results on the chemical composition of the Galactic ring nebula NGC~6888 surrounding the WN6(h) star WR136. The data are based on deep spectroscopical observations taken with the High Dispersion Spectrograph at the 8.2~\unit{m} Subaru Telescope. The spectra cover the optical range from 3700 to 7400 \AA. The effect of the CNO cycle is well identified in the abundances of He, N, and O, while elements not involved in the synthesis such as Ar, S, and Fe present values consistent with the solar vicinity and the ambient gas. The major achievement of this work is the first detection of the faint \ion{C}{2} \wav4267 recombination line in a Wolf-Rayet nebula. This allows to estimate the C abundance in NGC~6888 and therefore investigate for the first time the trace of the CNO cycle in a ring nebula around a Wolf-Rayet star. Although the detection of the \ion{C}{2} line has a low signal-to-noise ratio, the C abundance seems to be higher than the predictions of recent stellar evolution models of massive stars. The Ne abundance also show a puzzling pattern with an abundance of about 0.5~\unit{dex} lower than the solar vicinity, which may be related to the action of the NeNa cycle. Attending to the constraints imposed by the dynamical timescale and the He/H and N/O ratios of the nebula, the comparison with stellar evolution models indicates that the initial mass of the stellar progenitor of NGC~6888 is between 25~\unit{M_\odot} and 40~\unit{M_\odot}.   
\end{abstract}


\section{Introduction} \label{intro}   
Massive stars ($\geq20$~\unit{M_\odot}) play an important role in the chemical evolution and chemical enrichment of galaxies since they are major sources of metals in the Universe. They dominate the feedback to the local circumstellar environment during their relatively short lifetimes, between 3 and 10 million years \citep{Ekstrometal12}. Through the emission of ionizing photons, the action of their strong stellar winds and their death as core-collapse supernova, massive stars interact with their surroundings injecting great amounts of radiation, mass and mechanical energy. An obvious manifestation of these interactions is the production of ring nebulae, i.e. filamentary shells that are found around evolved massive stars, especially Wolf-Rayet (WR) stars \citep[e.g.][]{millerchu93, marstonetal94a, marstonetal94b, dopitaetal94, marston97, vanderhucht01}.  Actually, WR features revealing the interaction between massive stars and the interstellar medium (ISM) are often observed in the integrated spectrum of star-forming galaxies \citep[e.g.][]{Schaereretal99, LopezSanchezEsteban09, LopezSanchezEsteban10a}. 
\begin{figure*}
   \centering
   \includegraphics[width=15.5cm]{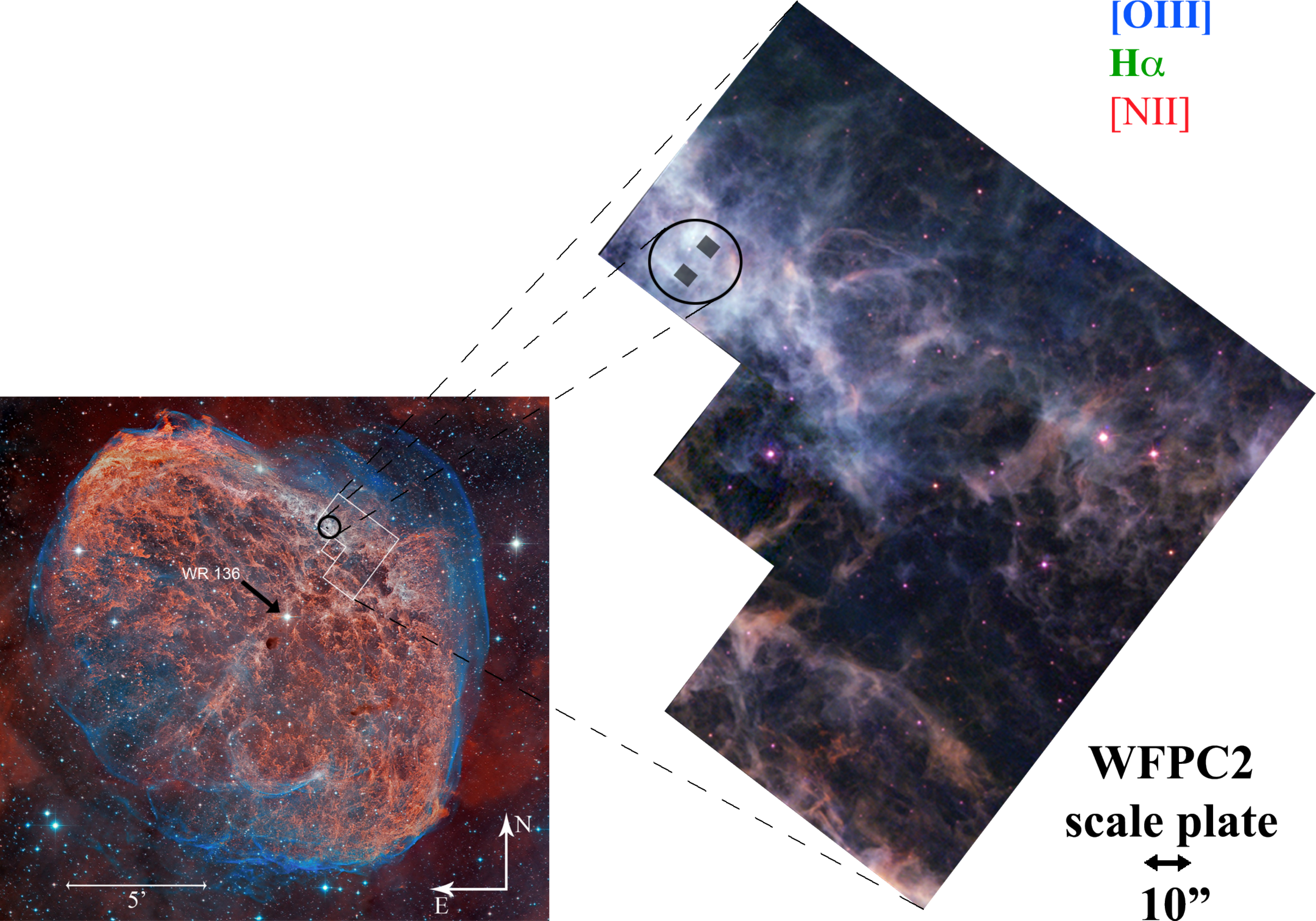} 
   \caption{The large extension of NGC~6888 is shown in the image obtained with the Wide Field Camera at the Isaac Newton Telescope (left, image release at http://www.ing.iac.es/PR/press/NGC6888.html). The image combines \halpha\ emission (red channel), 25\% \halpha\ and 75\% [\ion{O}{3}] emissions (green channel), and [\ion{O}{3}] emission (blue channel). The zoomed field to the right corresponds to an image observed with the Wide Field Planetary Camera 2 (WFPC2) on board the {\it Hubble Space Telescope} by the 8568 program. The image represents our own combination of the filters F502N ([\ion{O}{3}] \wav5007, blue), F656N (\halpha, green), and F658N ([\ion{N}{2}] \wav6583, red). The location of the slits observed with HDS are shown on the WFPC2 image. The slit covering the blue range is the one at the northwest and that corresponding to the red range is at the southeast. The slits were orientated with a position angle of 50$^\circ$ and cover an area of 3\arcsec$\times$4\farcs4.}
   \label{f1}
 \end{figure*}

The true nature of ring nebulae was first pointed out by \cite{johnsonhogg65} in the study of the Galactic objects NGC~6888 and NGC~2359. Today, their origin and evolution are understood in the framework of massive star evolution \citep[e.g.][]{GarciaSeguraetal96b, GarciaSeguraetal96a, Freyeretal03, Freyeretal06, van-Marleetal05, van-Marleetal07}. It is well established that ring nebulae are interstellar bubbles of ionized gas that have swept up the surrounding medium during the mass-loss episodes of their central massive stars \citep{weaveretal77}. The mass-loss episodes are simply the different phases of the stellar wind, which is especially powerful during the post-main sequence stages, playing an active role in the evolution of massive stars \citep{Vinketal01, Vinkjsetal11}. 

The interaction between the stellar wind and the circumstellar medium can produce ring nebulae with distinct characteristics. They will depend on the properties of the surroundings and of the stellar progenitor as well as its evolutive stage. Low-dense and hot interstellar bubbles are produced by the fast stellar winds of massive stars along the main sequence (MS) stage. They have been observed in the optical range \citep[e.g.][]{marston97, stockbarlow10}, at $21$~\unit{cm}  \cite[e.g.][]{Cappaetal05} and at $24$~\unit{\mu m} \cite[e.g.][]{Kraemeretal10}. These bubbles can be partially filled by the dense and slow stellar wind that operates during the phases of red supergiant (RSG) or luminous blue variable (LBV), depending on the initial mass of the progenitor. With the beginning of the WR phase, the material of the RSG wind is blown by the strong stellar wind and, finally, compressed into long filaments after the WR shell reaches the RSG shell. According to the calculations of \cite{van-Marleetal05}, the remaining shell will dissipate over $\sim80,000$ years after the beginning of the WR phase. In the literature, excellent works review in detail the evolutionary tracks of massive stars and the production of ring nebulae through the different stellar stages \citep[e.g.][]{Arthur07, Freyeretal06, toalaarthur11}. 

An interesting group of ring nebulae is that presenting chemical traces of processed stellar material;  the ejected and wind-blown types as defined in the classification of \cite{chu81}, recently reviewed by \cite{stockbarlow10}. These particular nebulae are thought to be blown by the stellar progenitors of long-duration $\gamma$-ray bursts \citep[e.g.][]{Vinketal11, Grafeneretal12}. They are powerful channels of information, directly connected to previous evolutionary phases. The enrichment process involves the transport of nucleosynthetic products from the stellar core to the outer layers, expelled in the post-MS stages. Thus, the study of their chemical abundances provides a unique opportunity to probe the stellar nucleosynthesis yield, an essential ingredient for our understanding of the chemical evolution of massive stars. From the analysis of He, N, O and Ne abundances, observational studies \citep[e.g.][]{estebanetal90, estebanetal92, stocketal11} have shown that ring nebulae containing stellar ejecta have overabundances of He and N, and a deficiency of O. The results indicate that a substantial fraction of O has been transformed into N in the stellar interior via the activation of the ON cycle. The enrichment pattern is consistent with the stellar nucleosynthesis of the H-burning through the CNO cycle, the main nuclear reactions working in massive stars along the MS. However, the lack of reliable determinations of C abundances in any ring nebulae has hindered for decades the full knowledge of the CNO cycle trace.

NGC~6888 is a prototype example of Galactic ring nebulae: a wind-blown bubble presenting clear traces of stellar nucleosynthetic products \citep[e.g.][]{kwitter81, estebanvilchez92}. A direct indication that NGC~6888 is a wind-blown bubble is the detection of X-ray emission in the bubble interior \citep[e.g.][]{Bochkarev88, Wriggeetal94}, produced by the interaction between the WR wind and the RSG material that heats up the gas at $\sim10^6$~\unit{K} \citep{weaveretal77}. Near the geometrical center of the nebula is located the stellar progenitor and ionizing source, the star HD~192163 \citep[WR136;][]{vanderhucht01}, a WN6(h)-type star in the WR phase with a mass of about 15~\unit{M_\odot} \citep{hamannetal06}. NGC~6888 is the best-studied nebula of its class thanks to its proximity \citep[$1.45\pm0.50$~\unit{kpc};][]{van-Leeuwen07} and its large extension of about $8\times5$~\unit{pc^2} ($18'\times12'$ on the sky). It resembles an almost complete ellipse, populated of filamentary structures of ionized gas. The mass of these structures amounts to 3.5--5~\unit{M_\odot} \citep{wendkeretal75, kwitter81, marstonmeaburn88}, with a typical electron density of about 300 \cmc\ \citep[e.g.][]{estebanvilchez92, fernandezmartinetal12}. The filaments mainly emit in [\ion{N}{2}] and \halpha, while they are surrounded by a highly-ionized, thin skin of material emitting in [\ion{O}{3}] \citep{mitra91, gruendletal00, mooreetal00}. The [\ion{O}{3}] layer represents the boundary between the MS bubble and the WR shell; instead, the filaments are RSG wind material, possible mixed with the WR wind, compressed after the WR shell collides with the RSG shell \citep[e.g.][]{kwitter81, estebanvilchez92, mooreetal00}. The gas kinematics have been studied in the past by several authors \citep[e.g.][]{chu88, marstonmeaburn88}, showing that NGC~6888 expands on average at 80 \kms, though the velocity field is complex with variations ranging from 55 to 110 \kms\ \citep{lozinskaya70}. From these velocities, the dynamical age of the WR shell is estimated in $\sim$30,000 years, varying between 20,000 and 40,000 years \citep[see][]{Freyeretal06}. 

In this paper, we review the CNO cycle trace in the chemical composition of the Galactic ring nebula NGC~6888, especially focusing on the determination of the total C abundance as derived from the faint \ion{C}{2} \wav4267 recombination line. In \S\ref{obsred} the observations are presented. A description of our spectrum as well as the methodology used in the line measurements and in the extinction correction can be found in \S\ref{med}. Physical conditions, chemical abundances and their calculations are found in \S\ref{results}. In \S\ref{chemo} the abundance pattern found in NGC~6888 is compared with those expected by the CNO cycle and the predictions by stellar evolution models. In \S\ref{discus}, the results are discussed in the light of previous ones for other similar objects and the future perspectives of exploring the C content in ring nebulae around massive stars. Finally, in \S\ref{conclu} we summarize our main conclusions. 

\section{Observations and data reduction} \label{obsred}  
NGC~6888 was observed on 2009 September 14 at Mauna Kea Observatory (Hawai'i), using the 8.2~\unit{m} Subaru Telescope with the High Dispersion Spectrograph \citep[HDS,][]{noguchietal02}. We used the blue and red cross dispersers with different inclination angles of the echelle grating in order to have an almost complete coverage of the optical range. The spectra taken with the blue cross disperser cover from 3667 to 5301 \AA\ with a gap between CCDs extending from 4436 to 4565 \AA. In the case of the spectra taken with the red cross disperser, they cover from 4777 to 7400 \AA\ with a gap between 5993 and 6237 \AA. For all spectra we used a slit length of 6$\arcsec$ and a width of 3$\arcsec$, leading to a spectral resolution of $R\sim12,000$. We take three consecutive exposures of 300 and 1800 seconds for the spectra obtained with the red and blue cross dispersers, respectively. These individual exposures were added to obtain the final spectra. The slit was located in a bright zone of the nebula located at the northwest of the ionizing WR star (see Fig.~\ref{f1}) and with a position angle of $50^{\circ}$. The atmospheric dispersion corrector (ADC) was used to keep the same observed region within the slit regardless of the air mass value. It was discovered while the data were being analyzed that the red and blue pointings were spatially offset ($\sim10"$; see Fig.~\ref{f1}), an effect we will discuss further in \S\ref{lmea}. Inspecting the headers of the data files we found that the mean coordinates of the three spectra of the blue range are $\alpha$ = 20$^{\rm h}$11$^{\rm m}$58\fs9 and $\delta$ =  +38$^\circ$24$^\prime$33\farcs3 (J2000.0); and of those of the red range:  $\alpha$ = 20$^{\rm h}$11$^{\rm m}$59\fs3 and $\delta$ =  +38$^\circ$24$^\prime$26\farcs8 (J2000.0). Unfortunately, this mismatch was unnoticed during the observations. 
 
The spectra were reduced using the {\sc iraf}\footnote{{\sc iraf} is distributed by NOAO, which is operated by AURA, under cooperative agreement with NSF.} echelle reduction package, following the standard procedure of bias subtraction, aperture extraction, flatfielding, background substraction, wavelength calibration, and flux calibration. A total area of 3\arcsec$\times$4\farcs4 was extracted to analyze the nebular emission. The standard stars BD~+28$^{\circ}$4211 and Feige~110 \citep{oke90} were observed to perform the flux calibration, whose uncertainty has been estimated to be of the order of 3$-$5\%. 
  \begin{figure}
   \centering
   \includegraphics[width=8.5cm]{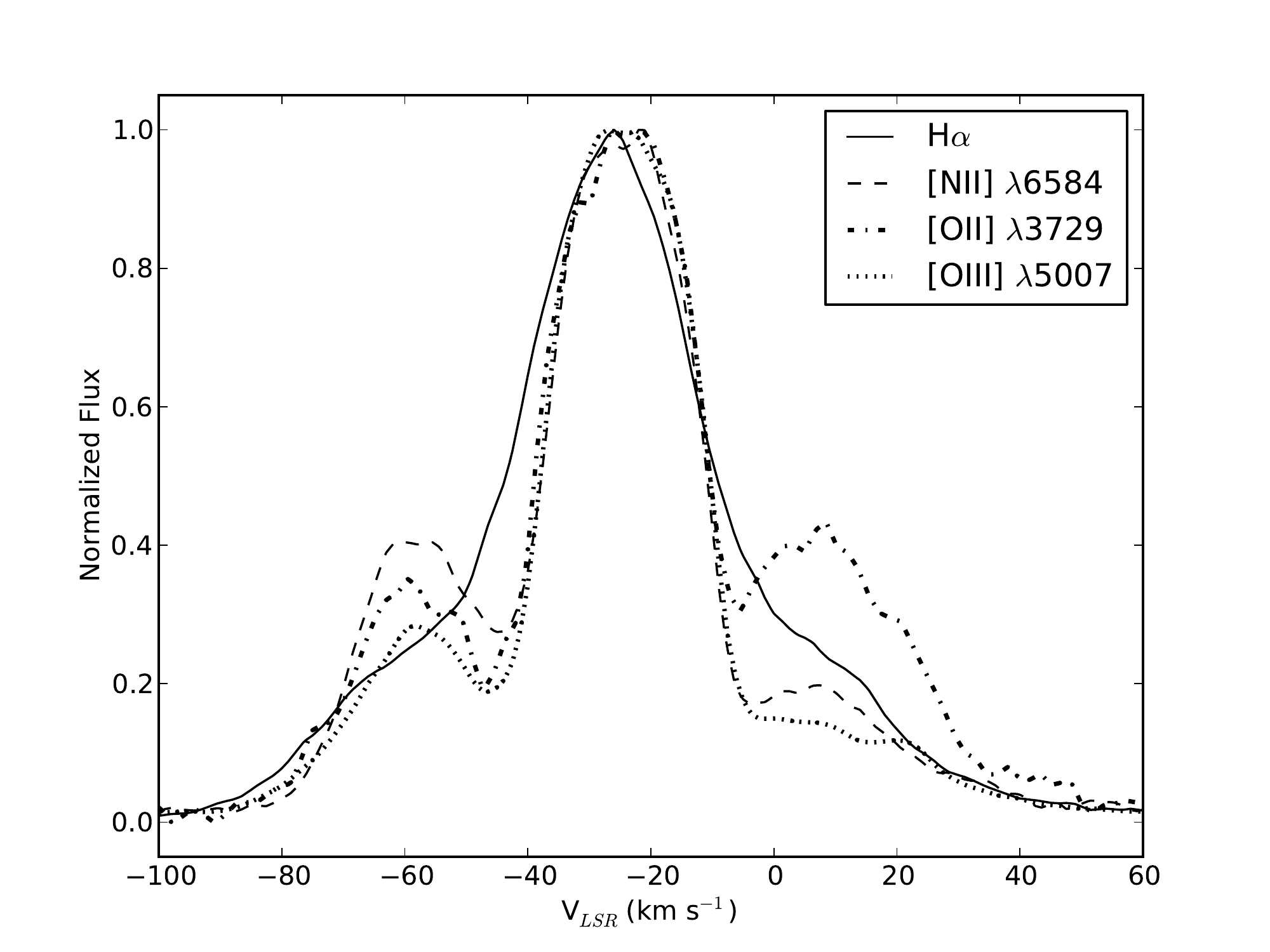} 
   \caption{Velocity profiles of selected emission lines: \halpha\ (solid line), [\ion{N}{2}] \wav6584 (dashed line), [\ion{O}{2}] \wav3729 (dotted-dashed line) and [\ion{O}{3}] \wav5007 (dotted line). Three kinematical components are detected at systemic velocities \vlsr\ of about $-$60, $-$25 and $+$12 \kms. }
   \label{vperfil}
  \end{figure}

\section{Emission lines and reddening correction} \label{med}
\subsection{Kinematical components} \label{kine}
The spectral resolution provided by HDS allowed us to detect three kinematical components in most emission lines. The velocity profiles of H$\alpha$, [\ion{N}{2}] \wav6584, [\ion{O}{2}] \wav3729 and [\ion{O}{3}] \wav5007 are presented in Fig.~\ref{vperfil}. The three components are clearly seen at systemic velocities \vlsr\ of about $-$60, $-$25 and $+$12 \kms. Though the red and blue ranges where observed at different positions on the northwest field, it should be noted that we detected the three kinematical components at the same velocities in both slit positions. 
The different width of the emission lines can be also seen in Fig.~\ref{vperfil}. {\halpha} is clearly wider due to its larger thermal broadening on account of its lower atomic weight.

In previous studies, the red-shifted component detected at $+$12 \kms\ in NGC~6888 has been usually associated with ambient emission from interstellar material, probably ionized by the Cygnus OB1 association \citep{lozinskaya70, trefferschu82, marstonmeaburn88, chu88}. The two blue-shifted components correspond to nebular material at different velocities in an approaching shell. Receding kinematical components of the shell were not detected in our spectra. According to \cite{trefferschu82}, that component is more prominent in the southwest part of the nebula. The absence of emission from the red-shifted component has been reported in previous works \citep{lozinskaya70, johnsonsongsathaporn81} and it can be associated with the presence of complex velocity structures near the edge of the shell, which is well populated of high-density knots (see Fig.~\ref{f1}). Using high-resolution Fabry-P\'erot observations with a resolution of 6.4 \kms, \cite{johnsonsongsathaporn81} mapped the [\ion{N}{2}] velocity structure in NGC~6888 finding the existence of three blue-shifted kinematical components in most of their positions at the edge of the nebula. Their positions 19 and 21 are very close to our slit positions with velocities \vlsr\ of about ($-$57, $-$31, $-$4), and ($-$35, $-$16, $-$1) \kms, respectively, which are in agreement with our determinations. Our blue-shifted components may correspond to unresolved blends of the velocity systems detected by these authors.

\subsection{Line measurements} \label{lmea}
Making use of the {\sc splot} routine of {\sc iraf}, line fluxes were measured applying a multiple Gaussian fitting over the average local continuum. Errors in the flux measurements were estimated from multiple measurements considering the continuum noise for all lines. The blend of \ion{He}{1} \wav3889 and H8 was difficult to resolve. Only the \ion{He}{1} and H8 emissions associated with the kinematical components at $-60$ and $+12$~\kms, respectively, were well isolated. We managed to estimate the fluxes of the three components of those lines from flux measurements of the blends and considering the theoretical H8/\hbeta\ ratio.

\begin{figure}
   \centering
   \includegraphics[width=8.5cm]{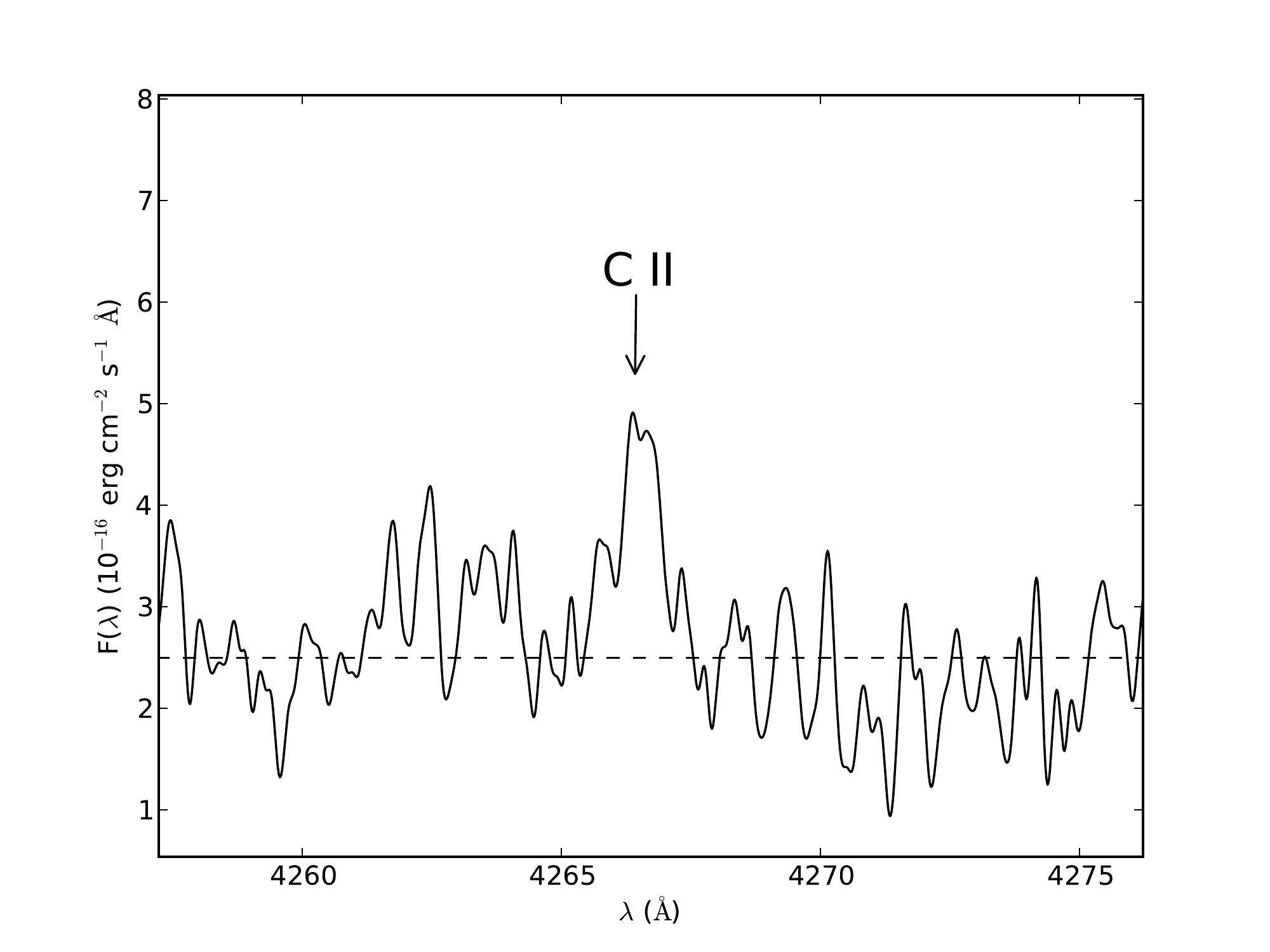} 
   \caption{Section of the echelle spectrum showing the \ion{C}{2} \wav4267 emission line, which is the first detection of this feature in NGC~6888. The dashed line corresponds to the average local continuum used to measure this line.}
   \label{c2line}
  \end{figure}

Integrated flux measurements were also considered for faint and single emission lines. In particular, this was the case of the recombination line \ion{C}{2} \wav4267, which represents the first detection of this feature in NGC~6888. The line was detected with a signal of 1.6$\sigma$ over the continuum. For a proper measurement of this line, a Gaussian filter was applied to smooth the noise of the local continuum resulting as it is shown in Fig.~\ref{c2line}. 

Three bright emission lines are presented in the common range of the blue and red spectra: \hbeta\ and [\ion{O}{3}] \wavs{4959}{5007}. We noticed a disagreement in the fluxes measured of 
those lines in both spectral ranges, thus confirming the pointing mismatch that we noted after the observations (see \S\ref{obsred} and Fig.~\ref{f1}). The flux measurements in red range were on average 34\% lower than those in the blue one for the three kinematical components.  Considering these differences, each spectral range was separately analyzed to produce a final homogeneous set of line flux ratios rescaled to its respective \hbeta. Errors in the flux ratios were propagated from the \hbeta\ normalization. 

\subsection{Reddening correction} \label{redcor}
Observed line fluxes, $F(\lambda)$, are affected by reddening due to the presence of dust along the line of sight. Given a reference emission line (typically, \hbeta), the reddening can be corrected through the usual expression:
\begin{equation}
 \frac{I(\lambda)}{I(H\beta)} = \frac{F(\lambda)}{F(H\beta)} 10^{c(H\beta)\times[f(\lambda) - f(H\beta)]}, 
\end{equation}
where $I(\lambda)$ represents the dereddened flux; the reddening coefficient, \chb, accounts for the amount of interstellar extinction at \hbeta; and $f(\lambda)$ is the adopted extinction curve. In our calculations, the reddening law $f(\lambda)$ of \cite{cardellietal89} was adopted using a ratio of total to selective extinction $R_V = 3.1$, the typical value of the diffuse interstellar medium. The extinction correction was calculated with {\sc pyneb} \citep{Luridianaetal12}, an updated {\sc python} version of the {\sc nebular} package of {\sc iraf}.
\begin{deluxetable*}{ccccccc}
\tabletypesize{\small}
\tablecaption{\label{flux}
Identifications and dereddened line ratios normalized to $I($\hbeta$)=100$ for the three kinematical components with velocities \vlsr\ of $-$60, $-$25, and $+$12 \kms\ as well as for the combination of the two blue-shifted components (Combined).}
\tablecolumns{7}
\tablewidth{0pc}
\tablehead{
\colhead{} & \colhead{}  & \colhead{}  & \multicolumn{4}{c}{$I(\lambda)/I($\hbeta)$^a$}\\
\cline{4-7}\\ 
\colhead{$\lambda$ (\AA)} & \colhead{Ion} & \colhead{Mult.} & \colhead{$-$60 \kms} & \colhead{$-$25 \kms} & \colhead{+12 \kms} & \colhead{Combined} 
}
\startdata
\multicolumn{7}{c}{Blue Range}\\
\hline
3726.03 &  [\ion{O}{2}] &  1F &  30$\pm$4 &  21$\pm$1 &  38$\pm$8 &  22$\pm$2 \\
3728.82 &  [\ion{O}{2}] &  1F &  33$\pm$5 &  23$\pm$1 &  53$\pm$10 &  24$\pm$2 \\
3750.15 &    \ion{H}{1} & H12 &   \nodata &   2.5$\pm$0.9 &   \nodata &	2.1$\pm$0.8 \\
3770.63 &    \ion{H}{1} & H11 &   \nodata &   4.1$\pm$0.5 &   \nodata &	3.5$\pm$0.5 \\
3797.63 &  [\ion{S}{3}] &  2F &   \nodata &    4.9$\pm$0.8 &   \nodata &	4.2$\pm$0.7 \\
3819.61 &   \ion{He}{1} &  22 &   \nodata &   2.1$\pm$0.6 &   \nodata &	1.8$\pm$0.5 \\
3835.39 &    \ion{H}{1} &  H9 &   7$\pm$2 &   7.9$\pm$0.6 &   10$\pm$3 &	7.8$\pm$0.8 \\
3868.75 & [\ion{Ne}{3}] &  1F &  22$\pm$3 &  16$\pm$1 &  17$\pm$7 &  17$\pm$2 \\
3888.65 &   \ion{He}{1} &   2 &  16$\pm$3 &   8.7$\pm$0.9 &   4$\pm$1 &	10$\pm$1 \\
3889.05 &    \ion{H}{1} &  H8 &  13$\pm$3 &  13$\pm$1 &  13$\pm$3 &  13$\pm$1 \\
3964.73 &   \ion{He}{1} &   5 &   \nodata &   0.8$\pm$0.3 &   \nodata &	0.7$\pm$0.2 \\
3967.46 & [\ion{Ne}{3}] &  1F &   3.9: &   4.2$\pm$0.6 &   4.0: &	4.1$\pm$0.6 \\
3970.07 &    \ion{H}{1} &  H7 &  14$\pm$3 &  15$\pm$1 &  14$\pm$5 &  15$\pm$1 \\
4026.21 &    \ion{He}{1} &  18 &   3.7: &   3.7$\pm$0.6 &   5.1: &	3.7$\pm$0.6 \\
4101.74 &    \ion{H}{1} &H$\delta$&  26$\pm$3 &  26$\pm$1 &  25$\pm$4 &  26$\pm$2 \\
4267.15 &    \ion{C}{2} &   6 &   \nodata &   0.8: &   \nodata &	0.7: \\
4340.47 &    \ion{H}{1} &H$\gamma$&  55$\pm$4 &  49$\pm$2 &  49$\pm$6 &  50$\pm$3 \\
4359.34 & [\ion{Fe}{2}] &  7F &   \nodata &   0.6$\pm$0.2 &   \nodata &	0.5$\pm$0.2 \\
4363.21 &  [\ion{O}{3}] &  2F &   1.5$\pm$0.6 &   1.5$\pm$0.5 &   \nodata &	1.5$\pm$0.5 \\
4387.93 &   \ion{He}{1} &  51 &   0.9$\pm$0.5 &   1.1$\pm$0.2 &   \nodata &	1.1$\pm$0.2 \\
4630.54 &    \ion{N}{2} &   5 &   \nodata &   0.6$\pm$0.2 &   \nodata &	0.5$\pm$0.2 \\
4658.10 & [\ion{Fe}{3}] &  3F &   2.3$\pm$0.8 &   1.2$\pm$0.3 &   \nodata &	1.4$\pm$0.3 \\
4713.14 &   \ion{He}{1} &  12 &   \nodata &   1.07: &   \nodata &	0.91: \\
4861.33 &    \ion{H}{1} &\hbeta& 100$\pm$3 & 100$\pm$1 & 100$\pm$3 & 100$\pm$1 \\
4921.93 &   \ion{He}{1} &  48 &   1.8: &   2.4$\pm$0.2 &   1.7: &	2.3$\pm$0.1 \\
4958.91 &  [\ion{O}{3}] &  1F & 106$\pm$3 &  91$\pm$2 &  49$\pm$2 &  93$\pm$2 \\
5006.84 &  [\ion{O}{3}] &  1F & 310$\pm$10 & 268$\pm$4 & 141$\pm$7 & 274$\pm$5 \\
5015.68 &   \ion{He}{1} &   4 &   4$\pm$1 &   4.5$\pm$0.1 &   3$\pm$1 &	4.4$\pm$0.2 \\
\multicolumn{3}{c}{$F$(\hbeta)$^b$} &  5.6$\pm$0.1 & 31.6$\pm$0.3 &  5.6$\pm$0.1 &  37.2$\pm$0.4 \\
\multicolumn{3}{c}{\chb} & 0.34$\pm$0.15 &  0.28$\pm$0.08 & 0.33$\pm$0.22 &  0.30$\pm$0.13\\
\hline
\multicolumn{7}{c}{Red Range}\\
\hline
4861.33 &    \ion{H}{1} &\hbeta& 100$\pm$7 & 100$\pm$1 & 100$\pm$7 & 100$\pm$1 \\
4958.91 &  [\ion{O}{3}] &  1F &  89$\pm$6 &  75$\pm$1 &  38$\pm$3 &  77$\pm$1 \\
5006.84 &  [\ion{O}{3}] &  1F & 254$\pm$18 & 218$\pm$3 & 113$\pm$8 & 224$\pm$3 \\
5875.64 &   \ion{He}{1} &  11 &  22$\pm$9 &  27$\pm$3 &  20$\pm$8 &  26$\pm$3 \\
6548.03 &  [\ion{N}{2}] &  1F &  60$\pm$6 &  41$\pm$1 &  29$\pm$3 &  44$\pm$1 \\
6562.82 &    \ion{H}{1} &\halpha& 285$\pm$29 & 285$\pm$6 & 285$\pm$29 & 285$\pm$6 \\
6583.41 &  [\ion{N}{2}] &  1F & 189$\pm$19 & 118$\pm$3 &  90$\pm$9 & 130$\pm$3 \\
6678.15 &   \ion{He}{1} &  46 &   \nodata &   8.2$\pm$0.5 &   \nodata &   6.8$\pm$0.4 \\
6716.47 &  [\ion{S}{2}] &  2F &   4.9: &   3.8: &  \nodata &   4.0: \\
6730.85 &  [\ion{S}{2}] &  2F &   \nodata &   4.6: &  \nodata &   3.8: \\
7065.28 &   \ion{He}{1} &  10 &   \nodata &   3.2$\pm$0.4 &   \nodata &   2.6$\pm$0.3 \\
7135.78 & [\ion{Ar}{3}] &  1F &  16$\pm$2 &  21$\pm$1 &   8$\pm$2 &  20$\pm$1 \\
\multicolumn{3}{c}{$F$(\hbeta)$^b$} &  2.0$\pm$0.1 & 10.6$\pm$0.1 &  2.1$\pm$0.1 &  12.6$\pm$0.1 \\
\multicolumn{3}{c}{\chb} & 0.43$\pm$0.10 &  0.32$\pm$0.03 & 0.35$\pm$0.10 &  0.34$\pm$0.03
\enddata
\tablenotetext{a}{Colons indicate an error equal or higher than 50\%}
\tablenotetext{b}{In units of $10^{-15}$ \unit{erg~cm^{-2}~s^{-1}} on an extracted area of 3\arcsec$\times$4\farcs4.}
\end{deluxetable*}

For each kinematical component, the extinction coefficient was determined by comparing the observed ratios of the available \ion{H}{1} emission lines relative to \hbeta\ with the theoretical ones for the case B predicted by \cite{storeyhummer95}, assuming an electron density \nel\ = 1000~\cmc\ and an electron temperature \te\ = 10,000~\K. Besides \hbeta, several \ion{H}{1} lines of the Balmer series were detected: \halpha, H$\gamma$, H$\delta$, H7, H8, H9, H11 and H12. In the red range, the \chb\ was directly given by the \halpha/\hbeta\ ratio. In the blue range, the adopted \chb\ was calculated as the weighted average of individual \chb\ values. In all cases, H8 was not considered since it is blended with \ion{He}{1} \wav3889. Also, H12 was discarded because its faintness only contributes to increase the error of the average extinction. H$\gamma$, H$\delta$, H7, H9, and H11 were used for both $-25$~\kms\ and combined blue-shifted components. In the fainter components, the \chb\ was calculated from the brighter lines: H$\delta$, H7, and H9 for the $-60$~\kms\ component; and H$\gamma$, H$\delta$, and H7 for the red-shifted component. H$\gamma$ was discarded in the $-60$~\kms\ component due to the reasons that we comment below. The final errors were obtained summing quadratically the weighted error and the standard deviation. In both spectral ranges, the different kinematical components present similar values of the reddening coefficient of about 0.3-0.4~\unit{dex} (see Table~{\ref{flux}}), in agreement with the typical ones found in other zones of the nebula \cite[e.g.][]{kwitter81, estebanvilchez92, fernandezmartinetal12}. 

During the extinction analysis, it was noted that the H$\gamma$/\hbeta\ ratio always returned a \chb\ coefficient lower than the other ratios, especially in the case of the $-60$~\kms\ component. The raw data do not show any anomalous effect around the H$\gamma$ emission. The flux calibration and the combination of orders were also checked, leading us to the same results. We concluded that this behavior might be related to radiative-transfer problems \citep{osterbrockferland06}. These effects would alter the \halpha/\hbeta\ and H$\gamma$/\hbeta\ ratios, returning discrepant reddening  values. Anomalous behaviors of the \chb\ coefficient determined from different pairs of Balmer lines have been reported and investigated in previous works \citep[e.g.][]{mesadelgadoetal09b}. Unfortunately, the dataset is not deep enough to explore this issue in NGC~6888 with the required accuracy.

In Table~\ref{flux}, we present  --for each spectral range-- the final list of line identifications (columns 1--3) as well as the dereddened flux ratios, $F($\hbeta), and \chb\ coefficients for the kinematical components $-$60\kms\ (column 4), $-$25 \kms\ (column 5) and $+$12 \kms\ (column 6). Additionally, we have included the line ratios associated with the sum of the two blue-shifted components, named as ``Combined" in column 7. For those lines only detected at $-$25 \kms, we just assumed that the measured flux comes from the two blue-shifted components. As we will see in \S\ref{results}, these two kinematical components have very similar physical and chemical properties and, therefore, it is convenient to analyze their sum. The errors associated with the line ratios include the uncertainties in the flux measurement, the normalization, and the reddening correction, which were propagated from equation (1).  

\section{Calculations and results} \label{results}
\subsection{Physical conditions} \label{phycon}
Physical conditions were calculated with {\sc pyneb} in combination with the atomic data listed in Table~\ref{atomic}. \te\ and \nel\ were determined making use of the diagnostic ratios available in our spectrum: [\ion{S}{2}] \wav6731/\wav6717 and [\ion{O}{2}] \wav3726/\wav3729 to derive \nel; and [\ion{O}{3}] \wav4363/\wav5007 to derive \te. All diagnostic lines were detected for the central component at $-25$~\kms. The [\ion{S}{2}] \wav6731 and [\ion{O}{3}] \wav4363 diagnostic lines were not detected in the kinematical components at $-60$~\kms\ and $+12$ \kms, respectively. Uncertainties in the physical conditions were propagated from the errors in the flux ratios.

The physical conditions are presented in Table~\ref{tphy}. As we already mentioned in \S\ref{redcor}, both blue-shifted components have very similar physical conditions and, therefore, the analysis of their combination also returns similar conditions. The densities in the blue-shifted components amount to 300 \cmc, while the red-shifted one has an \nel\ of about 70 \cmc. These determinations are consistent with results found in previous spectroscopic studies that show densities ranging from 100 to 600 \cmc\ in the nebula \citep[e.g.][]{kwitter81, mitra91, estebanvilchez92, fernandezmartinetal12} and a density of $<100$ \cmc\ in the ambient gas \citep{estebanvilchez92}. The \te\ derived from the [\ion{O}{3}] line ratio is of about 9000~\unit{K} in the blue-shifted components. This value is consistent with the analysis of \cite{kwitter81} and \cite{estebanvilchez92}, who found \te\ of about 8500-9500~\unit{K} in different areas of the nebula from the [\ion{N}{2}] and [\ion{O}{3}] diagnostic ratios. The results of \cite{mitra91} are the most discrepant, showing a \te\ of about 16,000~\unit{K} from the [\ion{O}{3}] line ratio in a zone near our slit positions. Very recently, \cite{fernandezmartinetal12} have carried out the first study of NGC~6888 using integral field spectroscopy. Those authors do not obtain direct determinations of 
 \te([\ion{O}{3}]), instead they estimate this quantity from direct determinations of \te([\ion{N}{2}]) and assuming a relation between both electron temperatures based on photoionization 
 models. In any case, three out of the nine apertures analyzed by  \cite{fernandezmartinetal12} show \te\ values of 8500-9000~\unit{K} that are similar to our determinations, while the \te\ is of about 6000-7000~\unit{K} in the other six apertures. 
\begin{deluxetable}{lcc}
\tabletypesize{\small}
\tablecaption{\label{atomic} Atomic dataset.}
\tablecolumns{3}
\tablewidth{0pc}
\tablehead{
\colhead{} & \colhead{Transition}  & \colhead{Collisional}\\
\colhead{Ion} & \colhead{Probabilities}  & \colhead{Strengths}
}
\startdata
N$^+$ & \cite{froesefischertachiev04} & \cite{tayal11} \\
O$^+$ & \cite{froesefischertachiev04} & \cite{kisieliusetal09}\\
\multirow{2}{*}{O$^{2+}$}&  \multirow{2}{*}{\cite{froesefischertachiev04}} & \cite{palayetal12}\\
               &  					        & \cite{aggarwalkeenan99}\\
Ne$^{2+}$& \cite{froesefischertachiev04} & \cite{McLaughlinBell00}\\
\multirow{2}{*}{S$^+$} & \cite{Podobedovaetal09} & \multirow{2}{*}{\cite{TayalZatsarinny10}}\\
           &  \cite{TayalZatsarinny10} & \\
Ar$^{2+}$& \cite{mendozazeippen83}& \cite{galavisetal95} \\
\multirow{2}{*}{Fe$^{2+}$}&  \cite{quinet96} & \multirow{2}{*}{\cite{zhang96}} \\
                 &\cite{Johanssonetal00}  &  
\enddata
\end{deluxetable}
\begin{deluxetable*}{cccccc}
\tabletypesize{\small}
\tablecaption{\label{tphy} Physical conditions$^a$.}
\tablecolumns{6}
\tablewidth{0pc}
\tablehead{
\multicolumn{2}{c}{Indicator} & \colhead{$-60$ \kms} & \colhead{$-25$ \kms} & \colhead{$+12$ \kms} & \colhead{Combined}
}
\startdata
\nel\ (\cmc)& [\ion{O}{2}] & 300:  & $310\pm30$ & 70: & $310\pm30$\\
                  & [\ion{S}{2}] & \nodata &1250:&  \nodata & 480:\\
\te\ (\unit{K}) & [\ion{O}{3}] & $8790\pm1140$ & $9100\pm810$ & $7300\pm400^b$ & $9050\pm810$
\enddata
\tablenotetext{a}{Colons indicate an error equal or higher than 50\%.}
\tablenotetext{b}{Fitting the O abundance to the Galactic O abundance gradient (see \S\ref{phycon}).}
\end{deluxetable*}
\begin{deluxetable*}{lccccc}
\tabletypesize{\small}
\tablecaption{\label{ionab} Ionic Abundances in units of 12$+log($X$^{+i}$/H$^+$).}
\tablecolumns{5}
\tablewidth{0pc}
\tablehead{
\colhead{Ion} & \colhead{Range} & \colhead{$-60$ \kms} & \colhead{$-25$ \kms} & \colhead{$+12$ \kms} & \colhead{Combined}
}
\startdata
He$^{+a}$ & Blue+Red& $11.16\pm0.09$ & $11.23\pm0.05$ & $11.09\pm0.12$ & $11.21\pm0.03$ \\
C$^{2+a}$& Blue& \nodata & $8.88\pm0.28$ & \nodata & $8.81\pm0.28$\\
N$^{+}$ & Red& $7.74\pm0.13$ & $7.51\pm0.08$ & $7.67\pm0.07$ & $7.54\pm0.08$\\
O$^+$& Blue& $7.64\pm0.23$ & $7.41\pm0.14$ & $8.20\pm0.12$ & $7.45\pm0.14$\\
O$^{2+}$& Blue& $8.22\pm0.17$ & $8.09\pm0.11$ & $8.22\pm0.12$ & $8.11\pm0.11$\\
& Red& $8.14\pm0.17$ & $8.01\pm0.11$& $8.12\pm0.08$ & $8.02\pm0.11$\\
Ne$^{2+}$& Blue& $7.49\pm0.27$ & $7.39\pm0.14$ & $7.86\pm0.18$ & $7.40\pm0.14$\\
S$^+$ & Red& $5.54\pm0.25$ & $5.44\pm0.12$ & \nodata & $5.43\pm0.09$\\
Ar$^{2+}$& Red& $6.30\pm0.17$ & $6.38\pm0.10$ & $6.24\pm0.11$ & $6.36\pm0.10$ \\
Fe$^{2+}$& Blue& $6.20\pm0.63$ & $5.85\pm0.28$ & \nodata & $5.92\pm0.30$
\enddata
\tablenotetext{a}{From RLs.}
\end{deluxetable*}

The absence of the [\ion{O}{3}] \wav4363 line made impossible to obtain an empirical determination of the \te\ for the red-shifted component. So, we followed an alternative approach. If the $+$12 \kms\ component is emitted by the ambient gas as previous studies have suggested \citep[e.g.][]{chu88}, the O/H ratio at the Galactocentric position of NGC~6888 would be given by the Galactic gradient of the O abundance (see \S\ref{cheab}). Then,  the \te\ of the ambient gas can be estimated fitting the O/H ratio predicted by the gradient with the O abundance derived from the observed [\ion{O}{3}] and [\ion{O}{2}] flux ratios of the blue range. The calculations return a \te\ of $7300\pm400$~\unit{K}, giving us a temperature estimation for the ambient gas in the Cygnus OB1 association. The uncertainty comes from the error propagation of the abundance gradient and the assumed distances (see \S\ref{cheab}). This temperature is consistent with the predictions of the Galactic gradient of \te([\ion{O}{3}]) derived by \cite{Deharvengetal00} from the analysis of Galactic \ion{H}{2} regions, $7350\pm500$~\unit{K}. The only previous estimation was done by \cite{estebanvilchez92}, who found a value of 6500~\unit{K} applying empirical flux--flux calibrations.

\subsection{Abundances from the empirical method}  \label{cheab}
Ionic abundances were computed from collisionally excited lines (CELs) using the {\sc pyneb} package and the atomic dataset of Table~\ref{atomic}.  The available ions in the blue spectra were O$^+$, O$^{2+}$, Ne$^{2+}$ and Fe$^{2+}$, while O$^{2+}$, N$^+$, S$^{+}$ and Ar$^{2+}$ were available in the red spectra. The physical conditions, \nel\ and \te, derived from the [\ion{O}{2}] and [\ion{O}{3}] diagnostic ratios in the blue spectra, respectively, were assumed for both spectral ranges. The final results are presented in Table~\ref{ionab}. The errors were estimated as the quadratic sum of the independent contributions of \nel, \te, and  flux ratio uncertainties.

Ionic abundances were also determined from recombination lines (RLs) for He$^+$ and C$^{2+}$. The adopted values are also presented in Table~\ref{ionab}. Currently, {\sc pyneb} provides optimal tools to calculate He$^+$ abundances. Several \ion{He}{1} lines were detected in the spectra of each component and slit position. We used the effective recombination coefficients of \cite{storeyhummer95} for \ion{H}{1} and those computed by \cite{porteretal12, porteretal13} for \ion{He}{1}, whose calculations include corrections for collisional excitation and self-absorption effects. The He$^+$/H$^+$ ratios for each kinematical component correspond to the weighted average values of the ionic abundances obtained from some individual lines: \ion{He}{1} \wav\wav4388, 4713, 4922, 5016, 5876, and 6678 for the combined and $-$25 km s$^{-1}$ components; \ion{He}{1} \wav\wav4388, 4922, 5016, and 5876 for the $-$60 km s$^{-1}$ component; and \ion{He}{1} \wav\wav4922, 5016, and 5876 for the +12 km s$^{-1}$ component. The associated errors in Table~\ref{ionab} represent the quadratic sum of the weighted error and the standard deviation.
\begin{deluxetable*}{lccccc}
\tabletypesize{\small}
\tablecaption{\label{icfs} Ionization correction factors.}
\tablecolumns{6}
\tablewidth{0pc}
\tablehead{
\colhead{Ratio} & ICF$^a$ & \colhead{$-60$ \kms} & \colhead{$-25$ \kms} & \colhead{$+12$ \kms} & \colhead{Combined} 
}
\startdata
He/H & He$^+ +$He$^{2+}$& $1.002\pm0.003$ & $1.001\pm0.002$ & $1.009\pm0.004$ & $1.001\pm0.002$ \\
C/O & C$^{2+}/$O$^{2+}$&\nodata & $0.93\pm0.01$ & \nodata  & $0.92\pm0.01$ \\ 
N/O$^b$&  N$^+$/O$^+$ & $1.68\pm0.29$ & $1.78\pm0.33$ & $1.34\pm0.14$ & $1.75\pm0.33$ \\
Ne/O & Ne$^{2+}$/O$^{2+}$& $1.05\pm0.06$ & $1.05\pm0.05$ & $1.00\pm0.01$ & $1.05\pm0.05$\\ 
S/O & S$^{+}$/O$^{+}$& $3.90\pm0.67$ & $3.94\pm0.63$ & \nodata & $3.94\pm0.63$\\ 
Ar/O & Ar$^{2+}$/(O$^{+}$+ O$^{2+}$) & $1.11\pm0.05$ & $1.13\pm0.07$ & $1.07\pm0.02$ & $1.13\pm0.06$ 
\enddata
\tablenotetext{a}{To be applied as: e.g.~N/O $=$ N$^+$/O$^+ \times$ ICF(N$^+$/O$^+$).}
\tablenotetext{b}{Assuming O$^+$/H$^+$ and O$^{++}$/H$^+$ from the blue range.}
\end{deluxetable*}

Our deep blue spectra allowed us to detect the faint \ion{C}{2} \wav4267 RL in the central kinematical component with a signal-to-noise ratio of 1.6 (see Fig.~\ref{c2line}). After correcting for the reddening, we obtained a flux ratio $I($\ion{C}{2})$/I($H$\beta) = 0.0081\pm0.0051$ in the $-$25~\kms component, which decreases to $0.0069\pm0.0043$ for the combined component. {\sc pyneb} does not support abundance calculations from RLs of heavy-element ions yet, so we implemented our own routines with the effective recombination coefficients of \cite{daveyetal00} to calculate the C$^{2+}$/H$^+$ ratios. The errors were also determined as in the ionic abundances from CELs, though the flux uncertainty dominates the error in this case.

From the analysis of the blue-shifted components and their combination, total abundances of He, C, N, O, Ne, S, Ar, and Fe were calculated for NGC~6888.   In \S\ref{chemo}, we present an extended discussion about the chemical composition of NGC~6888, while here we just focus on the calculations of the elemental abundances. The total O abundance was simply determined as the sum of O$^+$ and O$^{2+}$ in the blue range. Excluding O, the calculation of the total abundances of other elements required to correct for unseen ionization stages by means of ionization correction factors (ICFs). To calculate the total Fe abundances, the empirical fit of \cite{rodriguezrubin05} was considered, accounting for the absence of Fe$^+$ and Fe$^{3+}$. For He, C, N, Ne, S, and Ar, we considered the new set of ICFs recently developed by \citet[][in press]{DelgadoIngladaetal14} and computed for a large grid of photoionization models. To estimate the ICFs and their errors, we used the same method as described by these authors in their work: we extract a subset from the PNe\_2014 models stored in the 3MdB database provided by the authors, selecting from them all the models with solar abundances and an effective temperature of about $50,000$~\unit{K}. The temperature value corresponds to the effective temperature of WR136 \citep{estebanetal93, crowthersmith96}. Then, the ICFs were computed in terms of the O$^{2+}/($O$^+ + $O$^{2+})$ ratio and their values are presented in Table~\ref{icfs}. The total abundances from ions only observed in the red range were calculated adopting the O$^+$/H$^+$ and O$^{2+}$/H$^+$ ratios from the blue range --therefore assuming the same O abundance in both slit positions. This approximation is reasonable since both zones have a similar O$^{2+}$/H$^+$ ratio and probably similar ionization degrees. The total abundances are shown in Table~\ref{totab} , where we can note that the chemical abundances of the blue-shifted components are rather similar.
\begin{deluxetable}{lcccc}
\tabletypesize{\small}
\tablecaption{\label{totab} Total Abundances in units of 12$+log($X/H).}
\tablecolumns{5}
\tablewidth{0pc}
\tablehead{
\colhead{Element} & \colhead{$-60$ \kms} & \colhead{$-25$ \kms} & \colhead{$+12$ \kms} & \colhead{Combined} 
}
\startdata
He$^a$ & $11.16\pm0.09$ & $11.23\pm0.05$ & $11.09\pm0.12$ & $11.21\pm0.03$ \\
C$^a$ & \nodata & $8.93\pm0.31$ & $8.59\pm0.10^b$ & $8.86\pm0.31$ \\ 
N   & $8.64\pm0.31$ & $8.52\pm0.20$ & $8.11\pm0.16$ & $8.54\pm0.20$ \\
O   & $8.32\pm0.14$ & $8.17\pm0.09$ & $8.51\pm0.07^b$& $8.20\pm0.09$ \\
Ne & $7.61\pm0.35$ & $7.49\pm0.20$ & $8.15\pm0.21$ & $7.51\pm0.20$ \\
S & $6.80\pm0.38$ & $6.81\pm0.22$ & \nodata & $6.77\pm0.20$ \\
Ar & $6.34\pm0.17$ & $6.43\pm0.11$ & $6.27\pm0.11$ & $6.41\pm0.11$ \\
Fe &  $6.59\pm0.70$ & $6.26\pm0.34$ & \nodata & $6.33\pm0.36$
\enddata
\tablenotetext{a}{From RLs.}
\tablenotetext{b}{From the Galactic gradients of C and O (see \S\ref{cheab} for details).}
\end{deluxetable}

Table~\ref{totab} also contains the elemental abundances of He, C, N, O, Ne, and Ar in the ambient gas of the Cygnus OB1 association as determined from the $+12$~\kms\ component. S and Fe abundances were not derived due to emission lines associated with these elements were not detected in this component. These abundance calculations were possible taking into account that the O/H ratio is given by the Galactic O abundance gradient, which allowed us to derive the \te([\ion{O}{3}]) in this kinematical component (see \S\ref{phycon}). The adopted gradient was:
\begin{equation}
 12+log(O/H) = (8.869\pm0.049)-(0.045\pm0.005)\times R,
\end{equation}
which comes from the linear fit of O abundances from CELs in \ion{H}{2} regions \citep{estebanetal05, garciarojasesteban07} and is scaled to the Galactocentric distance of the Sun $R_0=8.0\pm0.5$~\unit{kpc} \citep{reid93}. To compute the O/H ratio, we estimated that the Galactocentric distance of NGC~6888 is $R=7.8\pm0.5$~\unit{kpc}, assuming a distance to NGC~6888 from the Sun of $1.45\pm0.50$~\unit{kpc} \citep[Hipparcos parallax to the central star WR136;][]{van-Leeuwen07}. Though the \ion{C}{2}~\wav{4267} RL was not detected in the red-shifted component, an estimate of its C/H ratio is useful to compare with the C abundance found in the ring nebula. We estimated that quantity from the Galactic C gradient of \cite{estebanetal05} that is based on the analysis of the \ion{C}{2} RL in \ion{H}{2} regions. He, N, Ne, and Ar abundances were calculated using the same ICF scheme presented above. The values of these ICFs are also presented in Table~\ref{icfs}. 

We have compared the total abundances of C, N, and Ne calculated above with alternative abundance determinations that involve other ICF schemes. The ICFs of C were obtained from the photoionization models of \cite{garnettetal99}, while N and Ne were derived using the ICFs proposed by \cite{peimbertcostero69}. The Ne/H ratio was also calculated with the ICF of  \cite{perezmonteroetal07}. In all cases, we found small differences within the estimated errors. 
\section{Chemodynamics of NGC~6888} \label{chemo}
The presence of stellar ejecta in ring nebulae allows us to better understand the nucleosynthesis of massive stars and the effects of the H-burning reactions on the elements involved in the CNO cycle. A detailed analysis of the chemical composition and abundance ratios of these elements provides us valuable information, which in turns can help us to constrain the evolutive scenario of the massive stellar progenitor of the ring nebula. In the following sections, we will review the final set of chemical abundances determined in this work for NGC~6888 and the surrounding medium as well as discuss their consequences on the CNO trace, and on the mass of the stellar progenitor when comparing with updated stellar evolution models. 
\subsection{Abundances and the CNO trace}  \label{checont}
In Table~\ref{comab} we summarize the final set of gaseous abundances of He, C, N, O, Ne, S, Ar, and Fe in the ambient gas of the Cygnus OB1 association and in the ring nebula NGC~6888. For comparison, we also include in Table~\ref{comab} the chemical composition of the solar vicinity represented by: the gaseous abundances of the Orion Nebula \citep{estebanetal04, rodriguezrubin05, simondiazstasinska11}; abundances in nearby early B-type stars \citep{NievaPrzybilla12}; and the solar ones considered in the stellar evolution models of \cite{Ekstrometal12}, who adopted the solar abundances ofÊ \cite{asplundetal05}, except for Ne which was taken from the analysis of nearby B-type stars of \cite{cunhaetal06}. In the second column of the table, we specified what kind of emission line was used in each abundance determination: CEL or RL. For clarification, we note that the information of this column only applies to the calculations of gaseous abundances and, thus, a direct comparison among the ambient gas, NGC~6888 and Orion is possible. 

A comparison with the B-star abundances or the adopted solar ones is also possible, but after correcting the gaseous abundances of heavy-elements by their depletion onto dust grains. C, O, and Fe are the only elements of Table~\ref{comab} where important depletions are expected. We have adopted the depletion values found in the solar neighborhood since NGC~6888 and the Cygnus OB1 association are only 1.45~\unit{kpc} away from the Sun. In the Orion Nebula, the O depletion is estimated to be $-0.12\pm0.03$~\unit{dex} \citep{mesadelgadoetal09b}, while the Fe depletion amounts to $-1.5\pm0.3$~\unit{dex} \citep{NievaPrzybilla12}. C is expected to be depleted in dust (PAH and graphite), though the study of its depletion is certainly problematic \citep[see][]{Mathis96a}. Following to \cite{estebanetal98}, we have adopted a C depletion of $-0.10$~\unit{dex}. It should be noted that these depletion factors might be actually different due to the effects of possible dust destruction processes associated with the strong gas flows present in NGC~6888.

Table~\ref{comab} also contains the elemental ratios of He, C, N, and Ne with respect to O for NGC~6888 and the ambient gas. The average ratios in the solar vicinity are also presented in Table~\ref{comab} as determined as the mean of Orion, B stars, and the solar metallicity. The errors of these values represent the standard deviation. To carry out a consistent comparison of the observed ratios with the average ones of the solar vicinity, two aspects have been considered in their calculations. On the one hand, the C and O abundances of the nebular objects were corrected by their depletion onto dust grains as we previously mentioned. On the other hand, the elemental ratios should be calculated from gaseous abundances that were determined by the same method, i.e$.$ using the same kind of emission lines (RLs or CELs); otherwise, ratios obtained through a combination of both emission lines can be severely affected by the so-called abundance discrepancy problem \citep[see][and references therein]{garciarojasesteban07}. Although RLs emitted by O ions were not detected in our spectroscopic analysis, we managed to estimate the O/H ratio from RLs in the ambient gas and NGC~6888 to properly calculate their corresponding He/O and C/O ratios. These abundance values are included in Table~\ref{comab}. For the ambient gas, the O abundance from RLs was obtained from the Galactic O gradient of \cite{estebanetal05} that is based on \ion{O}{2} RLs. In the case of NGC~6888, the abundance discrepancy problem itself gave us a way of estimating the O/H ratio from RLs. From the analysis of the largest sample of Galactic \ion{H}{2} regions, \cite{garciarojasesteban07} found that on average the RL-CEL discrepancy amounts to $0.16\pm0.06$ dex for O$^+$ and $0.25\pm0.02$ dex for O$^{2+}$. Then, the O/H ratio from RLs in NGC~6888 was estimated after adding the average discrepancies to the O$^+$ and O$^{2+}$ CEL-abundances derived through the empirical method. In Table~\ref{comab}, we can see that the RL-CEL discrepancy in the O abundance amounts to 0.23~\unit{dex} for NGC~6888. 

The comparison of the abundance ratios compiled in Table~\ref{comab} indicates the striking abundance pattern found in NGC~6888. Within the errors, the gaseous abundances of the ambient gas are basically similar to the Orion ones. The elemental ratios shows small differences when comparing with the solar average ratios, but consistent within the errors. These discrepancies may be produced by biases induced by the line de-blending process and/or ionization degree discrepancies because the pointing mismatch of the blue and red spectral ranges. In general, the agreement between the ambient gas and the solar neighborhood is expected considering the small variation rates of the Galactic gradients and the proximity of NGC~6888.

The chemical composition and elemental ratios of NGC~6888 show a clear enrichment pattern. The Fe abundance is similar to that of the Orion Nebula, which is expected considering that massive stars do not enrich the ISM with Fe prior to the supernova phase. This fact also indicates that eventual dust destruction does not seem to affect the gas-phase Fe abundance in a substantial manner. S and Ar also present abundances consistent with Orion since these elements are not affected by the CNO burning chains. Attending to the He, N and O abundances, we can distinguish the trace of the CNO cycle: He and N overabundances, and an O deficiency, produced by complete cycles of the CN and ON chains through the reactions $^{15}$N$(p, \alpha)^{12}$C and $^{17}$O$(p, \alpha)^{14}$N, respectively. Previous studies of NGC~6888 have also found a similar enrichment trace and total abundances of He, N and O consistent with our determinations. For example, \cite{estebanvilchez92} found He/H, N/H and O/H ratios of $11.30\pm0.09$, $8.40\pm0.35$ and $8.11\pm0.28$~\unit{dex}, respectively, from the analysis of a slit position near WR136. In a slit position near ours, \cite{mitra91} also found similar abundances of about 11.30, 8.14 and 8.15~\unit{dex} for He, N, and O, respectively. However, we should note that this slit position of \cite{mitra91} is where we found the largest discrepancy of temperatures with respect to our results as we mentioned in \S\ref{phycon}. On the other hand, the Ne/O ratio seems to be consistent with the solar values within the uncertainties, while the Ne/H ratio is clearly below solar. Attending to the C/H and C/O ratios, the comparison with the solar neighborhood values reveals an anomalous pattern. In \S\ref{cane}, we discuss in more detail these new findings in the framework of the CNO cycle and the nucleosynthesis of massive stars.  
\begin{deluxetable*}{lcccccc}
\tabletypesize{\small}
\tablecaption{\label{comab} Abundance comparison in units 12$+log($X/H).}
\tablecolumns{6}
\tablewidth{0pc}
\tablehead{
 & &  &  & \multicolumn{3}{c}{Solar Neighborhood$^b$}\\
Element & Method$^a$ & Ambient Gas & NGC~6888 &  Orion & B stars &  Solar
}
\startdata
He &RL & $11.09\pm0.12$ & $11.21\pm0.03$  & $10.991\pm0.003$ & $10.99\pm0.01$ & $10.93\pm0.01$ \\
C & RL & $8.59\pm0.10^d$ & $8.86\pm0.31$ & $8.37\pm0.03$ & $8.33\pm0.04$  & $8.39\pm0.05$ \\
N  & CEL &  $8.11\pm0.16$ & $8.54\pm0.20$ & $7.92\pm0.09$ & $7.79\pm0.04$ & $7.78\pm0.06$ \\
 \multirow{2}{*}{O} & CEL & $8.51\pm0.07$& $8.20\pm0.09$ & $8.52\pm0.01$ & \multirow{2}{*}{$8.76\pm0.05$} & \multirow{2}{*}{$8.66\pm0.05$}\\
 & RL &  $8.70\pm0.12^c$& $8.43\pm0.13^d$ & $8.65\pm0.03$ &  & \\
Ne &  CEL &$8.15\pm0.21$ & $7.51\pm0.20$ & $8.05\pm0.03$ & $8.09\pm0.05$ & $8.11\pm0.04$\\
S & CEL &  \nodata & $6.77\pm0.20$ &  $6.87\pm0.06$ & \nodata & \nodata \\
Ar & CEL &  $6.27\pm0.11$ & $6.41\pm0.11$&  $6.39\pm0.03$& \nodata &\nodata \\
Fe &  CEL &\nodata & $6.33\pm0.36$ & $6.00\pm0.30$ & $7.52\pm0.03$ & $7.45\pm0.05$ \\
\hline 
& & \multicolumn{2}{c}{Elemental Ratios$^e$} & \multicolumn{3}{c}{Solar Average Ratios$^f$}\\
\hline
He/O &  RL &$2.27\pm0.17$  & $2.66\pm0.13$ & \multicolumn{3}{c}{$2.24\pm0.03$}\\
C/O & RL & $-0.13\pm0.16$  & $0.41\pm0.34$   &  \multicolumn{3}{c}{$-0.33\pm0.08$}\\ 
N/O &  CEL &$-0.52\pm0.18$ & $0.22\pm0.22$ & \multicolumn{3}{c}{$-0.84\pm0.13$}\\
Ne/O &  CEL &$-0.48\pm0.22$ & $-0.81\pm0.22$  & \multicolumn{3}{c}{$-0.60\pm0.06$ / $-0.65\pm0.07^g$}
\enddata
\tablenotetext{a}{Only applies to the gaseous abundances calculated for the ambient gas, NGC~6888 and Orion.}
\tablenotetext{b}{Chemical content in the Solar vicinity (see \S\ref{checont}).}
\tablenotetext{c}{From the Galactic gradients based on RLs \citep{estebanetal05}.}
\tablenotetext{d}{Sum of the O$^+$ and O$^{2+}$ abundances from CELs after correcting them by the average RL-CEL discrepancy found in \ion{H}{2} regions (see \S\ref{checont}).}
\tablenotetext{e}{Corrected by C and O depletion onto dust grains (see \S\ref{checont}).}
\tablenotetext{f}{Average ratios of Orion (after correcting for C and O depletions), B stars and solar. Errors account for the standard deviation.}
\tablenotetext{g}{Computed adopting the Ne/H ratio for the Sun of $7.93\pm0.10$~\unit{dex} recommended by \cite{asplundetal09}.}
\end{deluxetable*}

\subsection{Chemical inhomogeneities in NGC~6888?}  \label{inho}
Up today, it is not completely clear that NGC~6888 is chemically inhomogeneous. A global analysis of the results from several slit positions distributed across the nebula \citep{kwitter81, mitra91} and the integral field study of \cite{fernandezmartinetal12} shows that He, N, and O abundances vary from 10.95 to 11.35~\unit{dex}, from 7.9 to 8.4~\unit{dex}, and from 8.1 and 8.6~\unit{dex}, respectively. However, it is necessary to emphasize that such a comprehensive comparison, mixing abundance determinations of different authors and years, should be regarded with caution. The use of different atomic data or unreliable detections of the \te-sensitive auroral lines are certainly important sources of error contributing to these variations. Comparing with Table~\ref{comab}, we see that the abundances we obtain are in the typical variation ranges. 

Recently, \cite{fernandezmartinetal12} has suggested the existence of chemical inhomogeneities in NGC~6888 and that they can be produced by material ejected by the central star at different moments along the post-MS stages. In favor of this hypothesis, \cite{Humphreys10} points out that mass-loss episodes can happen through non-isotropic and powerful ejections, which can explain the chemical inhomogeneities across a ring nebula associated with a massive star. This hypothesis is also plausible according to the nucleosynthesis predicted by stellar evolution models of massive stars. Figs.~\ref{ratios} and~\ref{ratios_40} show the time variation of different abundance ratios in the material ejected by 25 and 40~\unit{M_\odot} stars (see \S\ref{mass} for more details). If these temporal abundance variations are actually transformed into spatial abundance variations across NGC~6888, our assumption of considering the same O/H ratio in our two slit positions to derive the N abundance (see \S\ref{cheab}) might be inappropriate. However, taking into account that: (a) our positions are very close each other (less than 10\arcsec\ apart, see Fig.~\ref{f1}), and (b) the abundances obtained by \cite{mitra91} in a nearby position are consistent with our determinations, it seems reasonable to consider that NGC~6888 is chemically homogeneous at least in the area covered by our slit positions.
\subsection{Comparison with stellar evolution models} \label{mass}
The central star of NGC 6888, WR136, has a present-day mass of about 15~\unit{M_\odot} \citep{hamannetal06}, and has just entered the WR phase according to its classification as an early nitrogen-type WR star \citep[WN6(h);][]{vanderhucht01}. The detection of X-ray emission in NGC~6888 is an observational clue that support this idea \citep[e.g.][]{Wriggeetal94}, indicating that the WR wind is interacting with the previous RSG shell. Additionally, it has been estimated that NGC~6888 has a dynamical age of about $\sim$20,000-40,000 years (see \S\ref{intro}). This ring nebula is thus a relatively recent event in the evolution of WR136. Surely, the chemical evolution of the NGC~6888$+$WR136 system in the light of this single-star scenario could suffer important modifications if the presence of a binary companion is confirmed as previous works have suggested \citep[see][and references therein]{StevensHowarth99}. Since the binarity of WR136 is still an unresolved issue, hereinafter we will focus on the single-star case.

Under the above considerations, we aim to constrain the mass of the progenitor of the WR136 star combining the updated stellar evolution models of \cite{Ekstrometal12} and \cite{Georgyetal12} with the new set of gaseous abundances of He, C, N, O, and Ne determined in this paper. \cite{estebanvilchez92} were the first to attempt this question in NGC~6888 from their spectroscopical analysis in the optical range. Comparing the gaseous abundances of He, N, and O with the enrichment pattern predicted by the stellar evolution models of \cite{maeder90}, those authors concluded that NGC~6888 and WR136 were originated from a stellar progenitor with a mass between 25 and 40~\unit{M_\odot}. Since the early study of \cite{estebanvilchez92}, stellar evolutive models have been improved in several aspects, incorporating the effects of stellar rotation, new recipes to account for the mass-loss rates, or updating nuclear reaction rates \citep[see][and references therein]{Ekstrometal12}. Recently, \cite{toalaarthur11} have attempted a similar approach using updated stellar evolution models and the N/O ratio derived by \cite{estebanvilchez92}. However, these authors only focused on discussing the effects of stellar rotation on the observed abundance pattern and confirmed that the observed trace is consistent with the fact that WR136 is an early nitrogen-type WR star. 

\begin{figure}
   \centering
   \includegraphics[width=8.5cm]{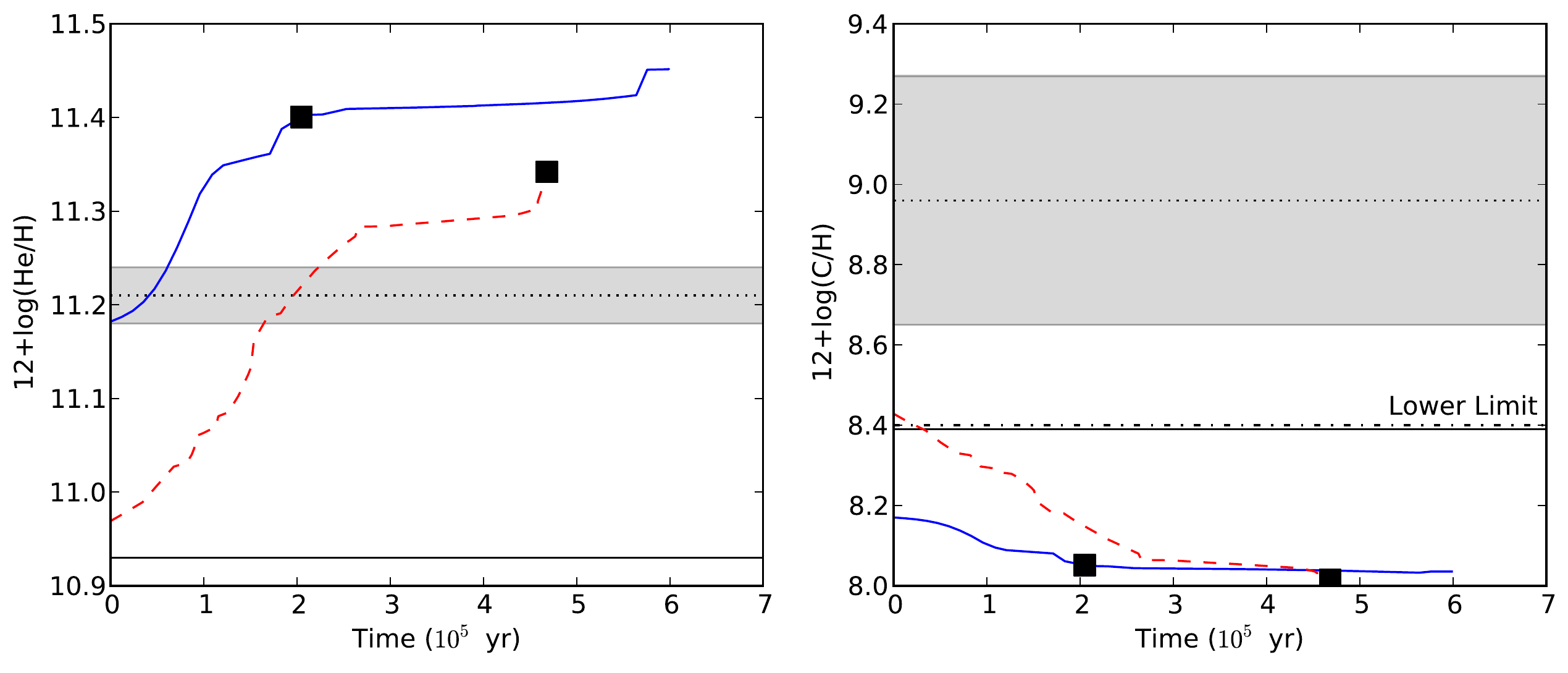} 
   \includegraphics[width=8.5cm]{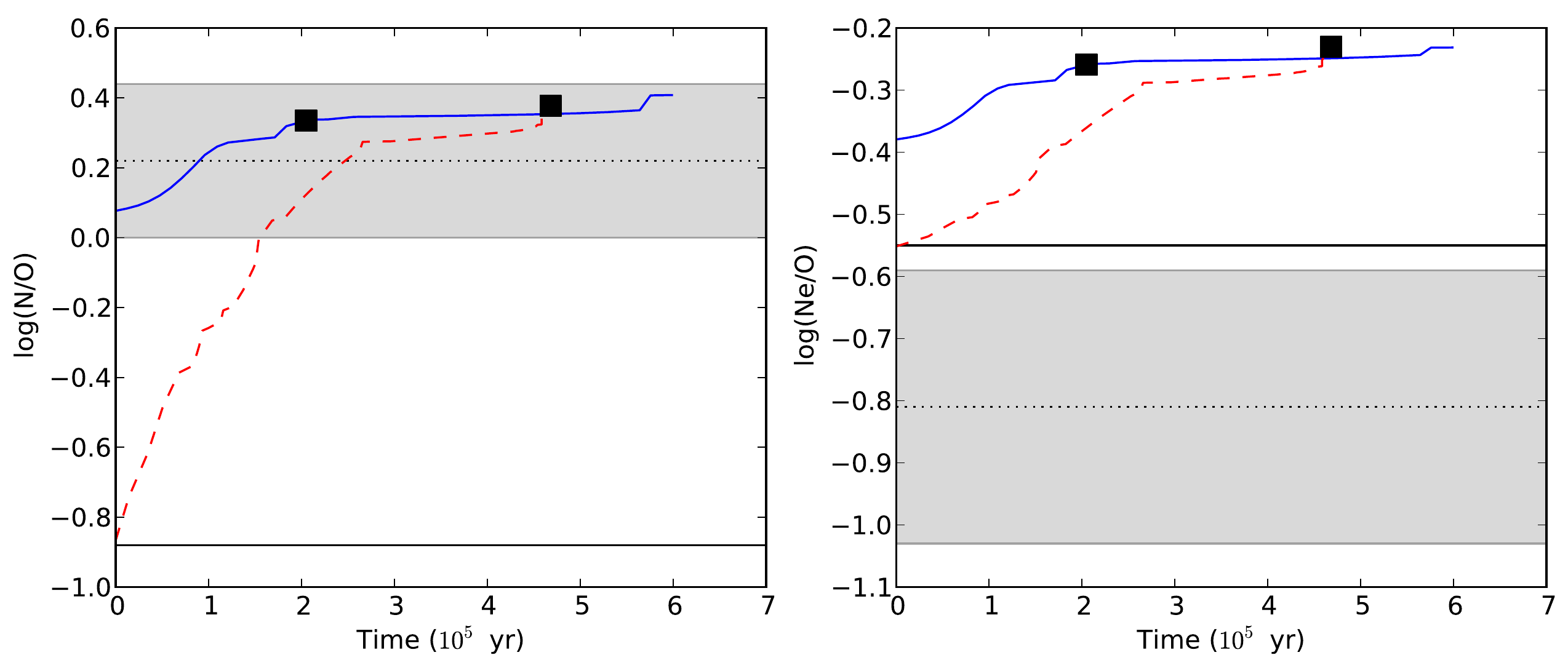} 
   \caption{He/H, C/H, N/O, and Ne/O abundance ratios of the stellar material ejected by a 25~\unit{M_\odot} star from the stellar evolution models of \cite{Ekstrometal12} and \cite{Georgyetal12} with rotation (blue solid line) and without rotation (red dashed line). The origin in abscissa corresponds to the onset of enhanced mass loss at the beginning of the RSG phase. The onset of the WR phase is indicated by the black square in both rotational and non-rotational stellar models. The horizontal solid lines correspond to the solar values as adopted by the stellar evolution models (see \S\ref{checont} for details). The grey bands represent the observed abundance ratios and their uncertainties. The dotted-dashed line almost overlapped with the solar one is the lower limit in the C/H ratio, which was estimated from the minimum value of C abundance compatible with the observational data (see \S\ref{mass} for details).}
   \label{ratios}
  \end{figure}

\begin{figure}
   \centering
   \includegraphics[width=8.5cm]{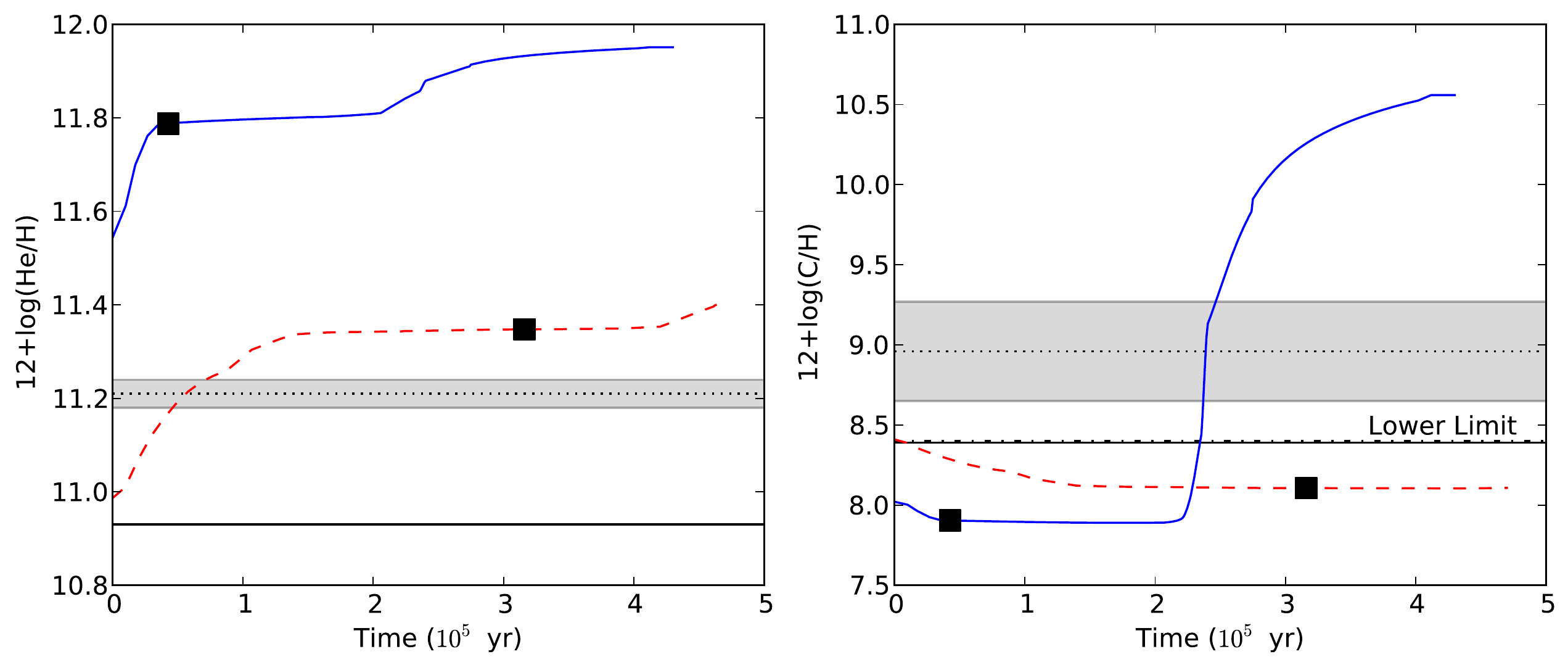} 
   \includegraphics[width=8.5cm]{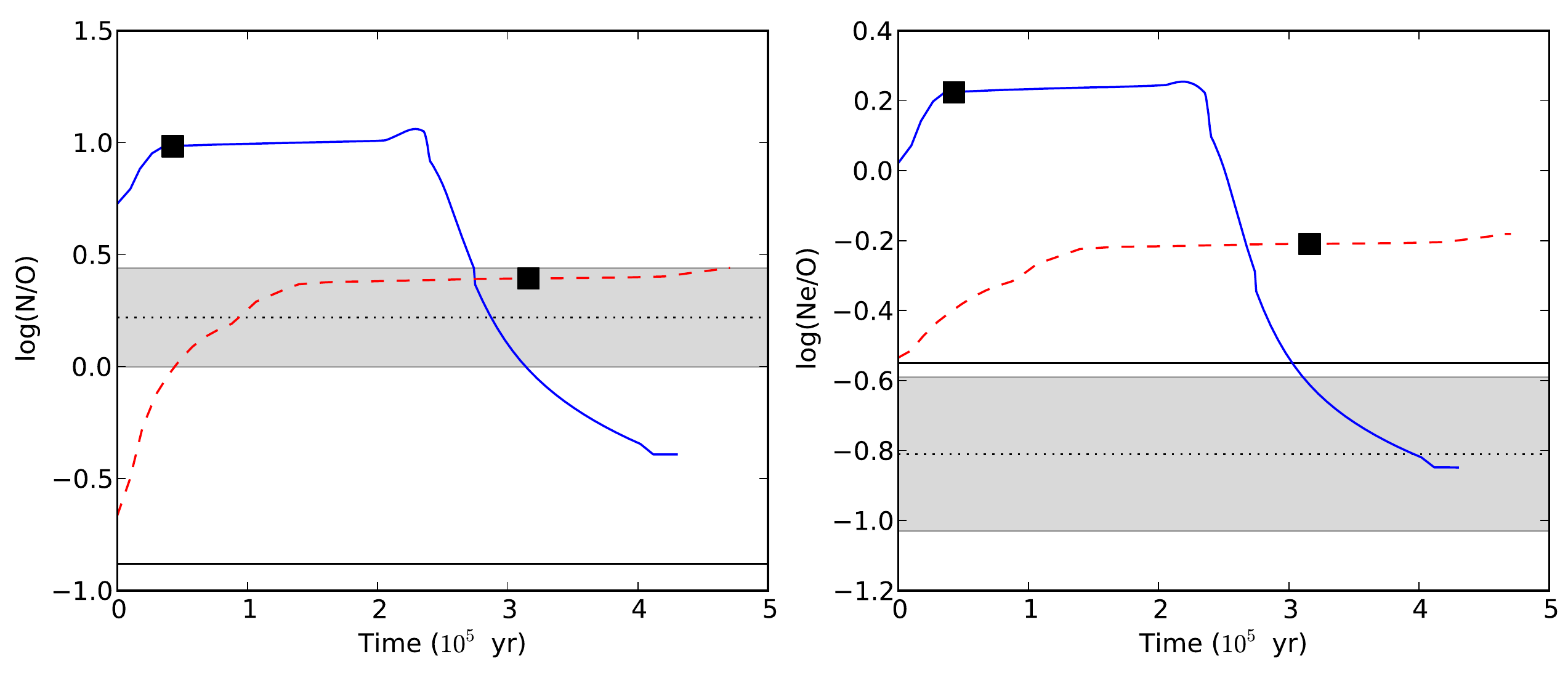} 
   \caption{Same as Fig.~\ref{ratios} for a 40~\unit{M_\odot} star.}
   \label{ratios_40}
  \end{figure}

In our analysis we have considered evolution models for stars with masses of 25, 32, 40 and 60~\unit{M_\odot}. To carry out the calculations, we have assumed that the stellar surface abundances given by the models are representative of the chemical composition of the stellar wind. The surface abundances of H, He, C, N, O, and Ne given by the models were weighted by the mass-loss rate and then integrated over time using the composite trapezoidal rule. The initial integration point was set in all models at the moment in which the massive star reaches the minimum effective temperature, $T_{eff}$, after finishing the MS stage. This point exactly coincides with the onset of enhanced mass loss at the beginning of the RSG phase (or LBV phase for the most massive stars) and, therefore, our approximation does not include material expelled during the MS stage. As \cite{toalaarthur11} argued, the contribution of the material ejected during the MS can be ignored due to it is spread throughout a much more extended bubble with a size of several tens of parsecs, while the RSG and WR wind material would form a ring nebula in the immediate surroundings around the central star. This last picture is consistent with the current view of the NGC~6888$+$WR136 system described in the first paragraph of this section. To properly compare the model predictions with the observed abundances, the integration results for He, C, N, O, and Ne relative to H and/or O were converted from fractional mass to number units through the atomic weight of the elements. For illustration, we present in Figs.~\ref{ratios} and ~\ref{ratios_40} the He/H, C/H, N/O, and Ne/O ratios for the stellar material ejected along time by a 25 and 40~\unit{M_\odot} star, respectively. We used the He/H and C/H ratios to explore the variation of the He and C abundances because these ratios are calculated from the intensity of RLs in both numerator and denominator. 
N and Ne abundances are determined from CELs and therefore we have used their ratios with respect to O abundances, which are also obtained from CELs. Using those particular abundance ratios, we avoid abundance discrepancy effects in the comparison.   

A quick look of Figs.~\ref{ratios} and~\ref{ratios_40} shows us the main differences between the rotational (dashed line) and non-rotational (solid line) models. It must be noted that rotational models are calculated assuming a very large initial rotation velocity of about 300 km s$^{-1}$, much larger than the present-day value of 37~\kms\ calculated by \cite{Grafeneretal12} for WR136. One of the most important effects of the stellar rotation is that the WR phase starts much earlier than what is predicted by the non-rotational models. Rotational models even predict that in the case of the 60~\unit{M_\odot} star the WR phase can be initiated during the MS stage. An interesting note about the new evolution models is also that the minimum mass for starting the WR phase is lower than in previous models, being now 20 and 25~\unit{M_\odot} for the rotational and non-rotational cases, respectively. In Figs.~\ref{ratios} and~\ref{ratios_40} the onset of the WR phase is indicated by squares. In all cases, this was set following the definitions of stellar types of \cite{Eldridgeetal08} and the considerations discussed by \cite{Georgyetal12}. The localization of this point is the first constraint that allows us to distinguish what evolution models are consistent with the current picture of the NGC~6888$+$WR136 system. According to the lifetimes of the WR phase estimated by \cite{Georgyetal12}, only the rotational models of 25 and 32~\unit{M_\odot}, and the rotational and non-rotational models of 40 and 60~\unit{M_\odot} are consistent with the dynamical age estimated for NGC~6888. 

In Figs.~\ref{ratios} and~\ref{ratios_40}, we can also see that the abundance patterns produced by both rotational and non-rotational models are very different. Qualitatively, the results for stellar models of 25, 32, and 40~\unit{M_\odot} are in general rather similar, though the timeline is certainly different. The C/H ratio presents the most distinctive pattern. Non-rotational models start the RSG phase with a C/H ratio similar to the solar value and rotational ones somewhat lower than solar. The C/H remains almost constant but, at a given moment, the enrichment pattern of the 25~\unit{M_\odot} stellar model behaves in a different way than those of the 32 and 40~\unit{M_\odot}. In both, the rotational and non-rotational cases of the 25~\unit{M_\odot} star, the C/H continues moderately decreasing until it reaches a saturation value that remains almost constant over time. The 32 and 40~\unit{M_\odot} stellar models show that: while non-rotational models keep a rather constant C/H ratio along time, rotational ones exhibit a remarkably strong C enrichment. This C is produced by $3\alpha$ reactions in the stellar interior, and transported to the stellar surface through efficient dredge-up and mixing processes as well as deep convective layers boosted by the stellar rotation. The abundance ratios of He/H, N/O, and Ne/O show a qualitatively similar behavior. Rotational models predict an important initial increase of He/H, N/O, and Ne/O at the beginning of the RSG phase, once again related to the effects of stellar rotation. Instead, non-rotational models clearly start the RSG phase ejecting material with solar abundance ratios. The case of the 60~\unit{M_\odot} stellar model produces abundance patterns with enhanced elemental ratios and slightly different behaviors than the models for stars of lower masses. However, we have not discussed it because this model is discarded as we will comment below.

We have compared the observed abundance ratios of NGC~6888 with the different model results. Given the current picture of the NGC~6888$+$WR136 system, the different ratios are associated with material that was probably ejected about one dynamical age of NGC~6888 ago or before the beginning of the WR phase. The observed ratios and their observational uncertainties are represented in Figs.~\ref{ratios} and~\ref{ratios_40} by means of grey bands over the predictions of the stellar evolution models.  The C/H and Ne/O ratios present puzzling patterns that do not fit the model predictions as we discuss in \S\ref{cane} and, therefore, our analysis is restricted to the He/H and N/O ratios. Attending to the time constraint of the dynamical age of the nebula and the He/H and N/O ratios, we found that our observations can be reproduced by the rotational model for a 25~\unit{M_\odot} star, and by the non-rotational models for 40 and 60~\unit{M_\odot} stars. The model of 60~\unit{M_\odot} can be rejected due to the inclusion of rotation in the new stellar evolution models predicts that the WR phase is initiated during the MS stage, which is not compatible with our current knowledge of WR136. Since rotation is certainly a property of every star in the Universe, our analysis suggests that NGC~6888 and WR136 are probably the evolutive products of a stellar progenitor with a mass lower than 40~\unit{M_\odot}. 

The comparison with stellar evolution models can also give an estimation of the time scale when the processed material was ejected by WR136. The non-rotational model of 60~\unit{M_\odot} predicts that the He/H and N/O ratios are representative of material ejected in different time scales before the onset of the WR phase. Considering the error bars, we obtained that the He/H ratio would represent material ejected about $15,000$~\unit{years} ago, while the material with the observed N/O ratio was expeled about $130,000$~\unit{years} ago. Instead, the rotational model of 25~\unit{M_\odot} and the non-rotational model of 40~\unit{M_\odot} produce more consistent results, where both observed ratios represents material ejected about $160,000$~\unit{years} and $260,000$~\unit{years}, respectively. If the observed material was ejected during the RSG phase and assuming that the RSG shell expanded with velocities similar to the RSG wind \cite[$10-20$~\kms;][]{van-Marleetal05}, we can estimate that the RSG shell has spent about $120,000-240,000$~\unit{years} to reach the location of our slit positions, about $2.5$~\unit{pc} away from the central star. This estimation is more consistent with the results found for 25 and 40~\unit{M_\odot} stars. 
\subsection{The puzzling C and Ne abundance patterns} \label{cane}
Unlike He and N, the comparison of the C abundance in NGC~6888 with the solar vicinity reveals at first look an enrichment pattern that does not fit the expectations of the CNO cycle (see Table 6 and \S\ref{checont}) and neither the C/H ratio can be reproduced by any stellar evolution model. In the case of the Ne abundance, we can see in Figs.~\ref{ratios} and~\ref{ratios_40} that the Ne/O ratio is always below the model predictions even considering the uncertainties. 

In the H-burning through the CNO cycle, C and O act as simple catalyzers driving the fusion reactions that convert $^1$H into $^4$He along the MS. However, given the different rates of the component reactions, the CNO cycle modifies the initial composition of the stellar material. The cycle efficiency is dominated by the slowest reaction, $^{14}$N$(p, \gamma)^{15}$O, responsible for the accumulation of N synthesized at the expense of C and O. We would then expect that the C/H ratio in NGC~6888 should be lower than the gaseous C abundance at the Galactic position of the nebula. At the Galactocentric position of the nebula, the C/H ratio of the ambient gas should range between 8.4 and 8.6~\unit{dex}, which are given by Orion and the Galactic C gradient, respectively. In contrast, we have estimated that the gaseous C abundance is about $8.9\pm0.3$ dex. We have estimated the strictest lower limit for the C/H ratio permitted by our data assuming that the continuum at the base of the \ion{C}{2} line is at the higher level of the noise estimated around the line. Following this procedure we obtain $I($\ion{C}{2} \wav4267)$/I($H$\beta)=0.0018$, which gives an abundance of 12+$log$(C/H)~$\simeq$~8.3. Assuming this value, the C/H ratio of NGC~6888 becomes similar to the expected one for its galactocentric position, but only considering non-rotational models for the 25, 32, and 40~\unit{M_\odot} at the beginning of the RSG phase. However, as we have already argued, rotation is an inherent property of every star and, therefore, this may be a rather improbable case. Indeed, rotational models predict C/H ratios lower than the estimated lower limit (see Figs.~\ref{ratios} and~\ref{ratios_40}).

If the C, N and O measured in NGC~6888 are produced by the CNO cycle, the sum of their abundances by mass should be similar to that of the ambient gas or, for example, the Orion Nebula. Calculating these numbers we find that using the nominal value of the 12+$log$(C/H) of 8.9~\unit{dex}, the sum of the CNO nuclei in NGC~6888 is a factor of two larger than expected and therefore an additional source of C should be necessary. This would imply that part of the freshly made C from the 3$\alpha$ reactions was  dredge-up to the stellar surface at the moment of the ejection of the bulk of the nebular material. On the other hand, assuming the strict lower limit of 12+$log$(C/H) $\sim$ 8.3 estimated above, the sum of CNO particles is in broad agreement with that expected by the CNO cycle, but implying that most of the N comes from the destruction of O, which is not a completely satisfactory solution. 

\cite{rolapelat94} argued that flux measurements of emission lines with low signal-to-noise ratio can be systematically overestimated, and such effect increases as the signal-to-noise ratio decreases. This effect may be affecting our C abundances, although some authors have advocated against it \citep[see][]{mathisliu99, esteban02}. Following the formulation of \cite{rolapelat94} provided to estimate such effect, we found that the measured flux of the \ion{C}{2} RL is reduced by around half. From the new flux ratio, we estimated a total C abundance of $8.6\pm0.3$~\unit{dex}. This value is certainly consistent with the gaseous C abundance of the ambient gas, but it still has the same issues than our lower limit (see previous paragraphs).

It is interesting to compare the C abundance determined here with the only previous estimation available in the literature for NGC~6888 obtained by \cite{mooreetal06}. Under geometrical assumptions and considering that most of the carbon is double-ionized, those authors estimated a C/H ratio of $7.6\pm0.3$~\unit{dex} from the analysis of the UV absorption line \ion{C}{3} \wav911 along the line of sight to the central star WR136. Correcting by C depletion onto dust grains, only the higher limit of this estimation is consistent with the results of the rotational stellar evolution models for all stellar masses. Obviously, there is a large discrepancy between the C/H ratio determined by \cite{mooreetal06} and us. This puzzling situation points out the real need of deriving more accurate C abundances in NGC~6888. 

Finally, as we already said in \S\ref{checont}, the Ne/O ratio of NGC~6888 seems to be solar within the uncertainties. However, in Figs.~\ref{ratios} and~\ref{ratios_40} it is clear that the Ne/O ratio is below the model predictions. In Table~\ref{comab} we can also see that the Ne/H ratio in the ambient gas and the solar vicinity are rather consistent, while NGC~6888 has a lower Ne abundance even within the uncertainties. The gaseous abundance of Ne in the Orion Nebula is indeed a reliable reference to compare with. From photoionization models with the data observed by \cite{estebanetal04}, \cite{simondiazstasinska11} derived a Ne/H ratio in Orion of $8.05$~\unit{dex} (quoted in Table~\ref{comab}), while \cite{rubinetal11} obtained a total Ne abundance of $8.00\pm0.03$~\unit{dex} from measurements of the infrared [\ion{Ne}{2}] and [\ion{Ne}{3}] CELs without requiring any ionization correction factor. Comparing with these values, the Ne/H in NGC~6888 is of about 0.5~\unit{dex} lower. To understand such behavior we should invoke the action of the NeNa cycle \citep[e.g.][]{ArnouldMowlavi93} in a higher intensity that the models assume. This cycle destroys Ne nuclei in the hydrogen burning zone at temperatures of about $3.5\times10^7$, values that can be reached in massive star cores. In fact, the action of NeNa cycle can be the reason of the large Na abundances found in yellow supergiants. \cite{denissenkov05} has proposed that rotational mixing of Na between the convective core and the radiative envelope in the MS progenitors of the more massive yellow supergiants can account for the Na enhancements observed. This would be a possible solution for the puzzling Ne underabundance observed in NGC~6888.  
\section{Carbon content in ring nebulae: a challenge for 10m telescopes and space observatories} \label{discus}

Until now, the chemical content of C in ring nebulae has been unknown and, therefore, we completely lacked on one of the fundamental parameters that constrains the enrichment trace of the CNO cycle. This has hindered for years a comprehensive comparison of the observed CNO trace with the predictions of nucleosynthesis models of massive stars. The determination of C abundances from the analysis of emission line spectra is in general a challenging task. In the particular case of ring nebulae, given their inherent low surface brightness ($F($H$\beta$)~$\leq 10^{-15}$~\unit{erg~s^{-1}~cm^{-2}~arcsec^{-2}}), it is certainly a challenge to overcome that needs the use of state-of-the-art technology in the largest aperture ground-based telescopes and space-borne observatories. 

Recently, \cite{stocketal11} have attempted without success the detection of [\ion{C}{1}] lines at 8727, 9824, and 9850 \AA\ in a sample of southern ring nebulae making use of the 3.6~m ESO New Technology Telescope and the 8~m Very Large Telescope. As the authors argued, these emission lines are usually observed in certain astrophysical contexts such as \ion{H}{2} regions or planetary nebulae. However, they may not be used as reliable empirical diagnostics since their emissions mainly arise behind the ionization front in a radiation-bounded region, produced by radiative transitions of neutral carbon previously excited by collision with free electrons. 

The detection of the brightest emission lines of carbon requires observations from space. On the one hand, the far-IR range contains the [\ion{C}{2}]~158 $\mu$m fine-structure line. Unfortunately, the use of this line is problematic because its emission predominantly comes from photodissociation regions \citep[e.g.][]{lebouteilleretal12}. This issue is also coupled with other problems related to IR spectroscopy such as ICFs, collisional de-excitation effects at modest electron densities, or still large sizes of the IR spectrograph beams \citep[see][]{garnettetal04}. The {\it Herschel} key problem MESS \citep[Mass-loss of Evolved StarS;][]{Groenewegenetal11} is the only project that today can provide crucial information about the C abundance in ring nebulae making use of the [\ion{C}{2}]~158 $\mu$m, but no results have come out in this sense yet. On the other hand, the UV range presents the well-known prominent carbon features \ion{C}{2}]~\wav2326 and \ion{C}{3}]~\wav1907$+$09, which are only observable by the {\it Hubble Space Telescope} (\hst). Unfortunately, the $HST$ have only accepted a single proposal (program ID: 8568) since its launch to investigate the C content in NGC~6888 using the Space Telescope Imaging Spectrograph and no C features were observed then. Since 2009 the unprecedented high-sensitivity of the Cosmic Origin Spectrograph at the $HST$ represents the only opportunity to detect the UV carbon lines, though we have still to take into account that UV observations can be severely affected by the uncertainties in the reddening correction and the strong dependence of the emissivity of UV CELs on the adopted electron temperature. Additionally, the \ion{C}{2}]~\wav2326 line is affected by similar concerns to those of the [\ion{C}{1}] and [\ion{C}{2}]~158 $\mu$m lines: given the low ionization potential of neutral carbon (11.2~\unit{eV}), it is expected the presence of C$^+$ behind the ionization front in a radiation-bounded region.

Thanks to new generations of CCDs with improved efficiency in the blue and the use of large aperture telescopes, determinations of the C/H ratio have become feasible with the detection of the \ion{C}{2} \wav4267 RL. Based on this RL, for example, it has been possible the study of radial gradients of C in the Galaxy and in a few external spiral galaxies \citep[e.g][]{estebanetal09, estebanetal13}. This emission line is however very faint, with a flux of about $0.001-0.01\times F($H$\beta$), requiring very deep observations for its detection. Up today, the \ion{C}{2} RL has only been detected in the Galactic ring nebulae NGC~6888 (this work) and NGC~7635 \citep{rodriguez99b, mooreetal02a, mesadelgadoesteban10}. In the particular case of NGC~7635, \cite{mesadelgadoesteban10} obtain 12+$log$(C/H) values between 8.6 and 9.0 in several areas of the nebula, abundances much larger than those expected from the Galactic abundance gradient, that should be about 8.45. As we can see, the behavior of C/H ratios in these two objects is rather similar and completely different to that shown by whatever \ion{H}{2} region. 

\cite{mesadelgadoesteban10} discussed the possible reasons of the high C/H 
ratios found in NGC~7635, and most of their arguments can be also applied for NGC~6888. Firstly, NGC~7635 is very unlikely that has suffered a pollution of C processed by the ionizing central star. It is an interstellar bubble blown by the young, massive central MS star BD$+60^o~2522$ of spectral type O6.5~IIIef \citep[e.g.][]{Dawanasetal07}. A mass of 45~\unit{M_\odot} has been estimated for BD$+60^o~2522$ \citep{HowarthPrinjal89, Dawanasetal07}, but it is thought that the star evolved from a progenitor of 60~\unit{M_\odot} \citep{Dawanasetal07}. \cite{christopoulouetal95} concluded that NGC~7635 nebula has a dynamical age of about $50,000$~\unit{years}. In contrast, NGC~6888 has been ejected by a post-MS star. In both cases, stellar evolution models do not predict a C enhancement 
in the evolution phase of the central stars when the nebular material was ejected. From the theoretical point of view, the only way to understand the strong C enrichment in the outer layers of the stellar progenitor of NGC~6888 --as discussed in \S\ref{mass}, is that the rotational-induced 
dredge-up of processed C takes place earlier in the evolution of the star, before the star enters the WR phase. Another possible explanation of a local increase of C may be due to the destruction of carbon-rich dust by the shocks associated with the expanding windblown bubble. This was already suggested by Moore et al. (2002) for NGC~7635. \cite{estebanetal98} estimated that the destruction of all the carbon locked-up onto dust grains in the Orion Nebula would increase the measured gas-phase C/H ratio by  only about 0.1 dex. An increase of such magnitude cannot explain completely the large C overabundances measured in the two ring nebulae with respect to the C/H ratios expected by the abundance gradients. A third explanation can be related to an abnormal enhancement in the intensity of the \ion{C}{2} \wav4267 RL. However, \cite{escalanteetal12} have recently concluded that \ion{C}{2} \wav4267 is not affected by 
any kind of fluorescence effect and therefore any possible mechanism that could affect the intensity of this line is actually unknown.

\section{Conclusions} \label{conclu}
We present results based on echelle spectroscopy of the Galactic ring nebula NGC~6888 taken with the High Dispersion Spectrograph at the 8.2~\unit{m} Subaru Telescope. The nebula is associated to the central star WR136, a massive Wolf-Rayet (WR) star of the WN6(h) sequence. The spectra are very deep and cover the optical range from 3700 to 7400 \AA. The high-spectral resolution of $R \sim 12,000$ permits to distinguish three kinematical components at $-$60, $-$25, and +12 km$^{-1}$. The two blue-shifted components correspond to gaseous filaments or clumps in the approaching side of the nebula, and the red-shifted one to the ambient gas that surrounds the Cygnus OB1 association. We have derived the physical conditions, \nel\ and \te, for each kinematical component. We obtain direct determinations of \te([\ion{O}{3}]) for the two blue-shifted components who belong to the shell of NGC~6888 itself, and an indirect estimation for the ambient gas. We determine ionic and total abundances of several elements from collisonally excited lines (CELs) as well as those of He and C from recombination lines (RLs). A comprehensive comparison of the gaseous abundances shows us that the effects of the CNO nucleosynthesis are well identified in He, N, and O, while elements such as Ar, S, and Fe --which are not involved in the cycle-- present abundances consistent with the solar values and the ambient gas. This is the first time that the \ion{C}{2} \wav4267 RL is detected in a ring nebula associated with a WR star, allowing to investigate the trace of the CNO cycle in a massive star ejecta. Although the detection of the \ion{C}{2} line has a low signal-to-noise ratio, the C abundance seems to be higher than the predictions of recent stellar evolution models of massive stars, even including the effects of stellar rotation. Further investigations are needed making use of large aperture telescopes to obtain more reliable measurements for this faint emission line. The Ne abundance also presents a puzzling pattern, being lower than the model predictions and lower than the solar vicinity. These Ne deficiency may be also related to the Na overabundances observed in yellow supergiants. The action of the NeNa cycle occurring in a higher rate than the models assume may explain these results.  Attending to the constraints imposed by the dynamical timescale of the nebula and the observed He/H and N/O ratios, the comparison with the predictions of stellar evolution models indicates that the initial mass of the stellar progenitor of NGC~6888 should be between 25 and 40 M$_\odot$.   

\acknowledgments
We are grateful to the referee, D. Stock, for his careful reading and proposed suggestions, which have help us to constrain the reliability of the carbon abundances. We thank to T.~H.~Puzia, J.~Arthur, G.~Meynet and, especially, M. Cervi\~no for their help with the evolutionary tracks. AMD thanks to R. Lachaume for being his {\sc python} guru. AMD acknowledges support from Comit\'e Mixto ESO-Chile, a Basal-CATA (PFB-06/2007) grant and the FONDECYT project 3140383. JGR and CEL acknowledge support from the Spanish Ministerio de Econom\'\i a y Competitividad (project AYA2011-22614). 

{\it Facilities:} \facility{SUBARU (HDS)}.


\end{document}